\documentclass[12pt]{article}
\pdfoutput=1
\usepackage{amsmath,amssymb,exscale,xcolor,multirow}
\usepackage[draft]{todonotes}
\usepackage{hyperref}
\input epsf
\newcommand{\m}[1]{\marginpar{{\tiny *}} }
\newcommand{\Dslash}{{\not \!\!D}}
\newcommand{\pslash}{{\not \!p\ }}

\begin{document}
\topmargin -1.0cm
\oddsidemargin -0.8cm
\evensidemargin -0.8cm
\begin{titlepage}
	
\begin{flushright}
	DESY 19-005
\end{flushright}
\thispagestyle{empty}
\vspace{40pt}

\begin{center}
\vspace{40pt}

\Large \textbf{The Minimal Simple Composite Higgs Model}

\end{center}
\renewcommand{\thefootnote}{\fnsymbol{footnote}}
\vspace{15pt}
\begin{center}
{\bf Leandro Da Rold$^{a,b}$\footnote{daroldl@cab.cnea.gov.ar}, Alejo N. Rossia$^{a,c}$\footnote{alejo.rossia@desy.de}} 
\vspace{20pt}

{\small\it
	$^a$ Centro At\'omico Bariloche, Instituto Balseiro\\Av. Bustillo 9500, R8402AGP, S. C. de Bariloche, Argentina\\[2mm]
	$^b$ CONICET, Av. de los Pioneros 2350, 8400 S. C. de Bariloche, Argentina\\[2mm]
	$^c$ DESY, Notkestra{\ss}e 85, D-22607 Hamburg, Germany \\[2mm]
	\par}

\end{center}

\vspace{20pt}
\begin{center}
\textbf{Abstract}
\end{center}
\vspace{5pt} {\small \noindent
Most of the analysis of composite Higgs have focussed on the Minimal Composite Higgs Model, based on the coset SO(5)$\times$U(1)$_X$/SO(4)$\times$U(1)$_X$. We consider a model based on the coset of simple groups SO(7)/SO(6), with SO(4)$\times$U(1)$_X$ embedded into SO(6). This extension of the minimal model leads to a new complex pNGB that has hypercharge and is a singlet of SU(2)$_L$, with properties mostly determined by the pattern of symmetry breaking and a mass of order TeV. Composite electroweak unification also leads to new bosonic and fermion resonances with exotic charges, not present in the minimal model, the lightest of these resonances being stable. A new rich phenomenology is expected at LHC.
}


\end{titlepage}
\noindent
\eject
\tableofcontents

\renewcommand{\thefootnote}{\arabic{footnote}}
\setcounter{footnote}{0}
\section{Introduction}
The hypothesis of the Higgs being a composite state offers one of the most interesting ideas to solve the hierarchy problem and explain the origin of electroweak symmetry breaking (EWSB). Constraints from new physics searches at Tevatron and LHC give bounds of order TeV on the scale $f$ of the new strongly interacting sector~\cite{Grojean:2013qca}. A separation between the EW scale and $f$ can be obtained if the Higgs is a pseudo Nambu-Goldstone boson (pNGB) of the new strongly interacting sector. The interactions with the elementary fermions of the Standard Model generate a potential at loop level, that can trigger EWSB dynamically. 

The nature of a strongly coupled field theory (SCFT) puts an unavoidable difficulty, given the lack of general non-perturbative methods that could allow to make precise predictions. With the advent of the AdS/CFT duality, extra-dimensional models were understood as weakly coupled holographic descriptions of four dimensional SCFTs~\cite{Maldacena:1997re}. Models defined in a slice of AdS$_5$ provided some of the most successful descriptions of this dynamics~\cite{Randall:1999ee}. Also discretised versions of extra-dimensions have been very useful for model building~\cite{ArkaniHamed:2001nc}.

The Minimal Composite Higgs Model (MCHM) is the realization of this idea with the minimal global symmetry group of the SCFT containing custodial symmetry and able to generate the Higgs as a pNGB~\cite{Agashe:2004rs}. The symmetry breaking pattern SO(5)$\to$SO(4) leads to only one SM multiplet of pNGBs, identified with the Higgs. The right normalization of hypercharge of the SM fermions required the introduction of an extra U(1)$_X$ factor in the EW sector. A significant amount of effort has been devoted, in the last decade, to the understanding of the physics of the MCHM. Also several extensions of the MCHM have been studied, like deformations of the extra-dimensional metric, as well as some extensions of the minimal group. Ref.~\cite{Gripaios:2009pe} has studied an extension to SO(6)/SO(5) that, in addition to the Higgs, contains an extra pNGB singlet, whereas Ref.~\cite{Mrazek:2011iu} has given a classification of cosets and has studied composite two-Higgs doublet models. Also interesting are the cases of composite grand unification with custodial symmetry, as in Refs.~\cite{Gripaios:2009dq,Frigerio:2011zg}.

In the present work we study a non-MCHM with composite EW unification. We embed the SO(5)$\times$U(1)$_X$ symmetry group of the MCHM into SO(7), with the spontaneous breaking SO(7)$\to$SO(6). Since SO(6) contains SO(4)$\times$U(1), identifying the abelian factor with U(1)$_X$ allows a unification of the symmetries. The coset SO(7)/SO(6) is the minimal one that generates the Higgs as a pNGB, contains the custodial symmetry and unifies SO(4)$\times$U(1)$_X$ into a simple Lie group. Besides the Higgs, it generates another pNGB that is an SU(2)$_L$-singlet and has non-vanishing hypercharge.

The SCFT produces fermion resonances transforming in irreducible representations of SO(7). We select a set of representations that allow mixing of the fermion resonances and the elementary fermions of the SM, and compute the one-loop potential based on symmetry considerations only. We analyse the conditions for suitable EWSB and the spectrum of the scalars. By using a 2-site description of the elementary-composite system, we are able to make numerical calculations of this potential, as well as the spectrum of bosons and fermions, finding regions of the parameter space with $f\sim 1.2$~TeV that can reproduce the SM spectrum. The mass of the new scalar is of order TeV.

The embedding of the EW symmetry into SO(7) gives a set of resonances with exotic charges, poorly explored in the literature. These resonances have a very rich phenomenology, as a lightest stable state that could be either the new pNGB or colored exotic fermions. The presence of these states can be problematic for cosmology, since they could lead to charged dark matter. However Refs.~\cite{DeLuca:2018mzn,Gross:2018zha} have shown that under suitable assumptions colored Dirac fermions with vanishing electric charge could be compatible with cosmology, and in some cases can even provide a fraction of dark matter.

The coset SO(7)/SO(6) has been considered previously in Refs.~\cite{Balkin:2017aep, Chala:2016ykx}, however the extra U(1) was not gauged there, thus composite EW unification was not achieved. Instead, the authors added an extra U(1)$_X$ group factor, as in the usual SO(5)/SO(4) models, leading to: SO(7)$\times$U(1)$_X$/SO(6)$\times$U(1)$_X$.
We remark the absence of that extra U(1)$_X$ factor in our setup, leading to a global symmetry completely described by a simple Lie group, up to the usual SU(3)$_c$ factor.
The gauging of the extra U(1)$\subset$SO(6) was considered in a different work~\cite{Balkin:2018tma}, but in their case the corresponding gauge field was a dark photon and the U(1)$_X$ needed for hypercharge was added as an extra factor, similarly to~\cite{Balkin:2017aep,Chala:2016ykx}. 
Since in those references the U(1)$\subset$SO(7) has no projection over hypercharge, the phenomenology of the resonances is different from the case that we consider. Another very interesting possibility is to include SU(3)$_c$ in the unified simple group, as in Refs.~\cite{Gripaios:2009dq,Frigerio:2011zg}. As we will discuss, the phenomenology of the present model contains new states with exotic charges, leading to a different phenomenology. 

The paper is organised as follows: in sec.~\ref{sec-model} we describe the symmetries of the model, the low-energy effective theory as well as the 2-site theory. In sec.~\ref{sec-potential} we compute the potential, study some approximations and calculate the spectrum of scalars. In sec.~\ref{sec-pheno} we show the spectrum of resonances, we analyse the phenomenology of the Higgs at LHC as well as its self-couplings and we give a brief description of the properties of the new pNGB state. We leave some discussions and conclusions for sec.~\ref{sec-conclusions}.

\section{A model from SO(7)/SO(6)}\label{sec-model}
In theories Beyond the Standard Model with a composite Higgs, the vacuum of the SCFT usually has a global unbroken symmetry SO(4)$\simeq$SU(2)$_L\times$SU(2)$_R$ that, after the Higgs acquires a vacuum expectation value (vev), preserves a custodial SO(3) symmetry that 
can protect the $\rho$-parameter. An extra U(1)$_X$ is required to obtain composite fermions with the same hypercharge as the SM quarks, with: 
\begin{equation}\label{eq-Y}
Y=T^3_R+\alpha X \ , 
\end{equation}
where $T^3_R$ is a generator of SU(2)$_R$ and $\alpha$ is a real constant. Therefore the vacuum of the SCFT has an unbroken semi-simple symmetry group SO(4)$\times$U(1)$_X$. In this paper, we will consider the case of a simple group, instead of a semi-simple one, concretely we will take the rank three group SO(6)$\supset$SO(4)$\times$U(1)$_X$. 

To obtain the Higgs as a pNGB, the SCFT will have a global symmetry group SO(7), spontaneously broken to SO(6) by the strong dynamics. The generators in the quotient SO(7)/SO(6) transform under SO(6) in the irreducible representation $\bf 6$, which under SO(4)$\times$U(1)$_X$ decomposes as: 
\begin{equation}\label{eq-rep-NGB}
{\bf 6}\sim({\bf2},{\bf2})_0+({\bf1},{\bf1})_{\pm 1/\sqrt{2}}\ . 
\end{equation}
Thus this pattern of symmetry breaking leads to a neutral bidoublet that can be identified with the Higgs, and a new state $\chi$ that is a singlet of SU(2)$_L$ and has non-vanishing hypercharge. 
We assume that the SCFT also has a global symmetry SU(3), that will be associated with the colour group of the SM.

The SCFT vacuum can be described by a vector $\Phi_0$, with the NGBs parameterising the fluctuations associated to the broken generators. By a suitable choice of basis (see App.~\ref{ap-so7}) we can write:
\begin{align}
&\Phi=e^{i\sqrt{2}\Pi/f}\Phi_0=\frac{\sin(\Gamma/f)}{\Gamma}(h_1,h_2,h_3,h_4,\Gamma\cot(\Gamma/f),\chi_1,\chi_2)\ ,\nonumber \\
&\Phi_0=(0,0,0,0,1,0,0) \ , \qquad \Gamma=\sum_{a=1}^6(\Pi^{\hat a})^2\ ,
\end{align}
where we have associated the first four broken generators with $H$ and the last two with $\chi$. $f$ is the decay constant of the NGBs.

A vev of $H$ or $\chi$ misaligns the vacuum, spontaneously breaking the EW symmetry of the SM. A vev of $\chi$ breaks U(1)$_X$, whereas a vev of $H$ breaks the EW symmetry as in the SM, leading to
\begin{equation}\label{eq-vev}
\Phi_v=(0,0,0,\sin(\langle h\rangle/f),\cos(\langle h\rangle/f),0,0) \ .
\end{equation}

Besides the NGB states, we assume that the SCFT also leads to vector resonances, that can be created by the Noether currents of the global symmetry, thus transforming in the adjoint representation of the corresponding groups. We also consider that the SCFT produces massive vector-like fermion resonances, that furnish full representations of the global symmetry. Unlike the case of the spin one resonances, the fermion representations are not fixed, leading to freedom for model building. We will study a particular choice of these representations, defined later in Eq.~(\ref{eq-ferm-rep}). We assume that the lowest level of resonances can be characterised by a single mass scale $m_\rho$, and the interactions between the resonances by a single coupling $g_\rho$, that will be taken as $g_\rho\gtrsim g_{\rm SM}$, but perturbative, and $f\sim m_\rho/g_\rho$. The masses and couplings of the different resonances do not need to be identical, but of the same order. 

The gauge and fermion fields of the SM are taken as external sources that probe the SCFT. A subgroup SU(2)$_L\times$U(1)$_Y$ of the SO(7) and the full SU(3) symmetry groups are weakly gauged by the SM. The elementary fermions interact linearly with the operators of the SCFT, leading to mixing with the resonances, and partial compositeness~\cite{Contino:2004vy}. As usual, we assume that the SCFT has an approximate scale invariance, such that the running of the linear coupling is controlled by the anomalous dimension of the SCFT operator. At low energies, if the scale at which the SCFT is defined is much larger than the TeV, a hierarchical pattern of flavor mixing can be generated, depending on whether the coupling is relevant, as needed for the top, or not, as needed for the first and second generations of fermions. Furthermore, we will assume that the SCFT is flavor anarchic~\cite{Agashe:2004cp}.

The elementary fields do not interact with the Higgs to leading order, such interactions are mediated by the mixing with the resonances that have the same quantum numbers as the elementary fields. The interactions of the elementary fields with the SCFT explicitly break the global symmetry, inducing a potential for the NGB at loop level. This potential is dominated by the contributions of the elementary fields with largest mixing, typically the Right-handed top and the Left-handed doublet of the third generation. The fermion contribution can misalign the vacuum and induce EWSB.

The vacuums $\Phi_0$ and $\Phi_v$ preserve a common SO(5) subgroup that contains the usual SO(3)-custodial group, as in the MCHM. This custodial symmetry protects the $\rho$-parameter. We will discuss fermion embeddings that also allow to protect some $Z$ couplings, as those of the Left-handed bottom quark and tau. The contributions to the $S$-parameter are similar to the MCHM, requiring a separation between $f$ and $\langle h\rangle$, roughly: $\sin^2(\langle h\rangle/f)\lesssim 0.1-0.2$~\cite{Grojean:2013qca}.

The scenario previously described can be realised by considering a theory with extra-dimensions, in particular in the presence of one compact extra-dimension with a metric that is AdS$_5$ near the UV boundary~\cite{Randall:1999ee,Contino:2003ve}. The elementary fields can be identified with the degrees of freedom of the UV boundary, and the SCFT with those of the bulk. It is also possible to give an effective description using a de-constructed version of the extra-dimensional theory, with a finite number of sites~\cite{ArkaniHamed:2001nc}. We will consider the simple case with two sites, one describing the elementary sector, and the other the first level of resonances~\cite{Contino:2006nn}, as well as a sigma model field connecting both sites. This description has more freedom than the extra-dimensional one, and has a cut-off not far from the scale of masses of the resonances. However it has shown to be very useful to parameterise this kind of theories and study their phenomenology at low energies and the LHC~\cite{Carena:2014ria}.

\subsection{2-site theory}\label{sec-2site}
We consider a site-0, also called elementary site, containing the same degrees of freedom as the SM, with the exception of the Higgs. We will denote the gauge couplings of site-0 generically as $g_0$. In order to reproduce the experimentally measured difference between $g$ and $g'$, we will also introduce a coupling $g_{0}'$ through the rescaling of the elementary part of the hypercharge gauge boson. It will be useful to work with an extension of the SM gauge symmetry group, therefore we define G$_0=$SO(7)$\times$SU(3). This extension can be realised at the level of fields by adding spurious degrees of freedom to furnish full representations of SO(7). These spurious fields are not dynamical, and they can be set to zero after the calculations~\cite{Agashe:2004rs}. 

On site-1, there is a gauge symmetry G$_1$=SO(7)$\times$SU(3), with gauge coupling generically denoted as $g_1\sim g_\rho$, that allows to introduce the lowest lying level of spin one resonances. In order to shorten notation, we will use small letters for fields at site-0, and capital letters for fields at site-1. 
The dynamics of the SCFT spontaneously breaks SO(7) to SO(6), generating a set of NGBs transforming in the fundamental representation of SO(6).
The NGB matrix is given by:
\begin{equation}
U_1=e^{i\sqrt{2}\Pi_1/f_1} \ , \qquad \Pi_1=\Pi_1^{\hat a}T^{\hat a} \ ,
\end{equation}
with $T^{\hat a}$ the broken generators of SO(7)/SO(6), $f_1$ the NGB decay constant and $\Pi_1^{\hat a}$ the NGB fields. Under SO(7), $U_1$ transforms as ${\cal G}U_1{\cal H}^\dagger$, with ${\cal G}\in$ SO(7) and ${\cal H}\in$ SO(6). 

The lowest dimensional representations of SO(7) of our interest are: the fundamental ${\bf7}$, the adjoint ${\bf21}$ and the ${\bf35}$. For these representations, $U$ can be expressed as:
\begin{equation}
U_1=I+i\frac{\sin(\Gamma_1/f_1)}{\Gamma_1}\Pi_1+2\frac{\cos(\Gamma_1/f_1)-1}{\Gamma_1^2}\Pi_1^2 \ ,\qquad \Gamma_1^2=\sum_{\hat a} (\Pi_1^{\hat a})^2 \ .
\end{equation}

There are also massive Dirac fermions in different representations of G$_1$, one for each SM fermion in a given representation of the SM gauge group. These Dirac fermions describe the lowest lying level of fermion resonances of the SCFT. For each generation of fermion resonances we choose the following representations of SO(7):
\begin{align}
&Q\sim {\bf21} \ , \qquad
U,D\sim {\bf35} \ , \nonumber\\
&L\sim {\bf7} \ , \qquad
E\sim {\bf21} \ ,
\label{eq-ferm-rep}
\end{align}
where $Q$ is the resonance associated to the quark doublet, $U$ and $D$ to the quark singlets, and $L$ and $E$ are the ones associated with the lepton doublet and singlet. Under SU(3) they transform in the same way as the associated elementary fermions. We have chosen all the generations transforming in the same representations of G$_1$, thus a generation index is understood in Eq.~(\ref{eq-ferm-rep}).

The aforementioned representations of SO(7) decompose under SO(6) as: 
\begin{equation}\label{eq-dec-76}
{\bf7}\sim{\bf6}+{\bf1}\ , \qquad
{\bf21}\sim{\bf15}+{\bf6}\ , \qquad
{\bf35}\sim{\bf15}+{\bf10}+\overline{\bf10}\ .
\end{equation}
To understand the transformation properties of the resonances under the custodial symmetry, we decompose the SO(6) representations under SO(4)$\times$U(1)$_X$: 
\begin{align}
&{\bf 10}\sim({\bf2},{\bf2})_0+({\bf3},{\bf1})_{1/\sqrt{2}}+({\bf1},{\bf3})_{-1/\sqrt{2}}\ ,\nonumber \\
&{\bf 15}\sim({\bf2},{\bf2})_{\pm 1/\sqrt{2}}+({\bf3},{\bf1})_0+({\bf1},{\bf3})_0+({\bf1},{\bf1})_0\ .
\label{eq-dec-64} 
\end{align}
The decomposition of ${\overline{\bf10}}$ can be obtained straightforwardly from the one of ${\bf10}$, and the decomposition of ${\bf6}$ is in Eq.~(\ref{eq-rep-NGB})

As is well known~\cite{Agashe:2006at}, one way to avoid large correction to the coupling $Zb_L\bar b_L$ in composite Higgs models, is to mix the elementary quark doublet $q_L$ with a fermion resonance with $T^3_R=-1/2$. This choice fixes the value of $\alpha$ in Eq.~(\ref{eq-Y}) to:
\begin{equation}\label{eq-Yalpha}
\alpha=\frac{2\sqrt{2}}{3} \ .
\end{equation}
With this choice, the components with $T^3_R=-1/2$ of the $({\bf2},{\bf2})_{\pm 1/\sqrt{2}}$ contained in the ${\bf15}$ of SO(6) have the same charges under the SM gauge symmetry as $q_L$. The corresponding partners of $u_R$ and $d_R$ are contained in the $({\bf1},{\bf3})_{1/\sqrt{2}}$ of $\overline{\bf10}$, with $T^3_R=0,-1$ respectively. For the leptons, the $\ell_L$ partners are the components of $({\bf2},{\bf2})_0$ with $T^3_R=-1/2$, contained in the ${\bf6}$, whereas the singlet is the component of $({\bf1},{\bf3})_0$ with $T^3_R=-1$, contained in the ${\bf15}$. A summary of the transformation properties of the resonances that can mix with the elementary fermions is shown in Table~\ref{t-ferm}.
\begin{table}[tb]
\begin{center}
\begin{tabular}{|c|c|c|c|c|}
\hline
\rule{0mm}{5mm}
Field & $T^3_R$ & SO(4)$\times$U(1)$_X$ & SO(6) & SO(7) \\ [0.3em]
\hline
\rule{0mm}{5mm}
$q$ & -1/2 & $({\bf2},{\bf2})_{1/\sqrt{2}}$ & ${\bf15}$ & ${\bf21}$ \\ [0.3em]
\hline
\rule{0mm}{5mm}
$u$ & 0 & $({\bf1},{\bf3})_{1/\sqrt{2}}$ & $\overline{\bf10}$ & ${\bf35}$ \\ [0.3em]
\hline
\rule{0mm}{5mm}
$d$ & -1 & $({\bf1},{\bf3})_{1/\sqrt{2}}$ & $\overline{\bf10}$ & ${\bf35}$ \\ [0.3em]
\hline
\rule{0mm}{5mm}
$\ell$ & -1/2 & $({\bf2},{\bf2})_0$ & ${\bf6}$ & ${\bf7}$ \\ [0.3em]
\hline
\rule{0mm}{5mm}
$e$ & -1 & $({\bf1},{\bf3})_0$ & ${\bf15}$ & ${\bf21}$ \\ [0.3em]
\hline
\end{tabular}
\end{center}
\caption{Transformation properties and embedding of elementary fermions into the representations of the fermion resonances they mix with.}
\label{t-ferm}
\end{table}

By making use of the transformation properties of $U$, it is possible to build SO(7)-invariants that superficially look like SO(6)-invariants only. These invariants allow us to introduce interactions between the fermion resonances and the NGBs. 

The Lagrangian of site-1 is:
\begin{align}
{\cal L}_1=&-\frac{1}{4g_1^2}F^a_{\mu\nu}F^{a,\mu\nu}+\frac{f_1^2}{4}d^{\hat a}_\mu d^{\hat a,\mu}+\bar Q(\Dslash-m_Q)Q +\bar U(\Dslash-m_U)U+\bar D(\Dslash-m_D)D\nonumber\\
&+\bar L(\Dslash-m_L)L+\bar E(\Dslash-m_E)E +f_1y_{U}[(\bar Q_L U_1)_{\bf 15}(U_1^\dagger U_R)_{\bf 15}]_{\bf1} \nonumber\\
&+f_1y_{D}[(\bar Q_L U_1)_{\bf 15}(U_1^\dagger D_R)_{\bf 15}]_{\bf1}
+f_1y_{E}[(\bar L_L U_1)_{\bf 6}(U_1^\dagger E_R)_{\bf 6}]_{\bf1} +{\rm h.c.}
\label{eq-L1}
\end{align}
$F^a_{\mu\nu}$ is the field strength of the gauge fields at site-1 (a sum over the terms of the gauge of SO(7) and SU(3)$_c$ is understood), $m_\Psi$ are the masses of the fermion resonances, taken of order $g_1f_1$, and $y_{\Psi}$ are dimensionless couplings of order $g_1$. Given the properties of $U_1$,  $U_1^\dagger \Psi$ transforms as a reducible representation of SO(6) under transformations of SO(7), thus $(U_1^\dagger \Psi)_{\bf r}$ is the projection on the representation ${\bf r}$ of SO(6). For the choices in Eq.~(\ref{eq-ferm-rep}), the quarks share the SO(6) representation $\bf 15$, whereas the leptons share the $\bf 6$. For other embeddings, one has to decompose the SO(7) representations under SO(6), and for the Yukawa interactions one has to project and sum over the common SO(6) representations. $d_\mu^{\hat a}$ is obtained from the Maurer-Cartan form, according to: $U_1^\dagger D_\mu U_1=ie_\mu^aT^a+id_\mu^{\hat a}T^{\hat a}$, thus the second term of Eq.~(\ref{eq-L1}) contains the kinetic term of the NGBs as well as interactions.~\footnote{We have not included four-fermion operators, they will be induced by exchange of spin one resonances.}

The last three terms of Eq.~(\ref{eq-L1}) include interactions among the pNGBs and fermion resonances as well as mixing terms among the latter. In particular, the first non-trivial order in its expansion in powers of the pNGBs fields is formed by Yukawa interactions, therefore these terms are generically called composite Yukawa terms. We have not included the most general kind of composite Yukawa terms to preserve the finiteness of the one-loop potential of the pNGBs \cite{Carena:2014ria}. The terms included have the same chiral structure as the SM counterparts. 

Site-0 and site-1 are connected by $\sigma$-model fields $\Omega=e^{i\sqrt{2}\Pi_0/f_0}$, transforming in the bifundamental of G$_0\times$G$_1$: $\Omega\to{\cal G}_0\Omega{\cal G}_1^\dagger$, with one field $\Omega$ for SO(7) and another one for SU(3). $\Pi_0$ parameterises the coset G$_0\times$G$_1$/G$_{0+1}$ and $f_0$ is the decay constant of these NGBs, taken of the same order as $f_1$. These fields provide the longitudinal polarization of the spin one resonances. The Lagrangian connecting both sites is:
\begin{align}
&{\cal L}_{\rm mix}=\frac{f_0^2}{4}|D_\mu\Omega|^2+f_0\sum_{i}\lambda_i\bar\psi_i\Omega\Psi_i+{\rm h.c.} \ , \nonumber\\
& \psi_i=q,u,d,\ell,e \ , \qquad \Psi_i=Q,U,D,L,E \ ,
\label{eq-Lmix}
\end{align}
where $\lambda_i$ is a mixing parameter, that determines the degree of compositeness of the SM fermions. 

We assume flavor anarchy of the composite sector, meaning that all the coefficients of the Yukawa couplings of the resonances are of the same order, as well as their masses. In this case the mixings can be fixed requiring a hierarchical spectrum for the SM fermions and a suitable CKM matrix. We consider this is the case in the present work, following Refs.~\cite{Agashe:2004cp,Csaki:2008zd,Agashe:2008uz}. Thus the mixings are hierarchical, being of order $g_1$ for $q_L$ and $u_R$ of the third generation, and much smaller for the other fermions. These are the only small parameters of the 2-site theory.

\subsubsection{Mass eigenstates}
${\cal L}_{\rm mix}$ provides masses for the fields in the coset G$_0\times$G$_1$/G$_{0+1}$, whereas the fields in G$_{0+1}$ remain massless, leading to the SM gauge couplings:
\begin{equation}\label{eq-matchingg}
\frac{1}{g^2}=\frac{1}{g_{0}^2}+\frac{1}{g_{1}^2} \ ,
\qquad
\frac{1}{{g'}^2}=\frac{17}{9}\left(\frac{1}{{g_0'}^2}+\frac{1}{g_{1}^2}\right) \ .
\end{equation}
where $g_0$ and $g_0'$ are the gauge couplings of SU(2)$_L$ and U(1)$_Y$ at site-0, and $g_1$ is the gauge coupling of SO(7) at site-1. A matching similar to the one of $g$ is present for the coupling of SU(3).

By a suitable gauge transformation, the fields $\Pi_0$ become the longitudinal degrees of freedom of the massive vector fields, and the physical NGB can be parameterised by
\begin{equation}
U=e^{i\sqrt{2}\Pi/f} \ , \qquad \Pi=\Pi^{\hat a}T^{\hat a} \ ,
\qquad
\frac{1}{f^2}=\frac{1}{f_0^2}+\frac{1}{f_1^2} \ .
\end{equation}
Among the 6 physical NGBs we find the 4 degrees of freedom which are identified with the SM Higgs boson. The remaining two degrees of freedom conform the additional charged scalar $\chi$ with electric charge $2/3$.

The site-1 contains 21 massive resonances of spin one. When the gauge coupling constants of the site-0 elementary gauge fields are set to zero, the resonances associated to the generators in SO(7)/SO(6) have mass $g_1\sqrt{(f_{0}^{2}+f_{1}^{2})/2}$ and the rest have mass $g_{1}f_{0}/\sqrt{2}\equiv m_\rho$. 

Setting the  gauge coupling constants of the site-0 to a non-zero value makes five of the latter resonances to mix with those elementary gauge fields of the site-0 which have the same transformation properties. The result of such mixing is three vector resonances which transform like the generators of SU(2)$_L$ with mass $f_{0}\sqrt{(g_{0}^2+g_{1}^{2})/2}$, one vector resonance which transforms like the hypercharge generator with mass $f_{0}\sqrt{({g_0'}^2+g_{1}^{2})/2}$. Those four resonances are partially composite states. There is one more resonance which is fully composite, has mass $g_{1}f_{0}/\sqrt{2}$ and is associated to the linear combination of $T_{3}^{R}$ and $X$ orthogonal to $Y$. 

Finally, the mixing gives rise to four partially composite massless vector bosons which are identified with the SM electroweak gauge bosons, $W_{\mu}^{i}$ and $B_{\mu}$, that can be written as linear combination of the elementary and composite gauge bosons:
\begin{align}
W_{\mu}^{i}= &\cos\left(\varphi\right)w_{\mu}^{i}+\sin\left(\varphi\right)A_{\mu}^{L,i} \label{e:SMWboson}\\
B_{\mu}=&\cos\left(\omega\right)b_{\mu}+\sin\left(\omega\right)\left[\cos\left(\theta_Y\right)A_{\mu}^{R,3}+\sin\left(\theta_Y\right)A_{\mu}^{X}\right],
\label{e:SMBboson}
\end{align}
where we have defined the elementary-composite mixing angles $\varphi$ and $\omega$ as:
\begin{equation}
\tan\left(\varphi\right)=\frac{g_{0}}{g_{1}},\qquad \tan\left(\omega\right)=\frac{{g_0'}}{g_{1}}.
\end{equation}
And the angle $\theta_Y$, which measures the mixing between $T_{3}^{R}$ and $X$ to give the $Y$ operator, is defined such that:
\begin{equation}
\tan\left(\theta_Y\right)=\alpha=\frac{2\sqrt{2}}{3}.
\end{equation}
The states in Eqs. (\ref{e:SMWboson}) and (\ref{e:SMBboson}) will acquire a non-zero mass only when the Higgs boson gets a non-zero vev, just like in the SM. EWSB will also induce additional mixings among states with equal electric charge, but those mixings are expected to be small.

The site-0 contains massless chiral fermions with identical transformation properties as the SM fermions. On the other hand, in the site-1 there are several vector-like fermion resonances. Each multiplet of resonances has a diagonal mass matrix. The values of those masses are independent but they will be considered to be $\mathcal{O}\left(\text{TeV}\right)$.

When considering the site-1 in isolation, not all the fermion resonances will have the mass given by the associated diagonal mass matrix because the terms in the second line of Eq.~(\ref{eq-L1}) generate mixing among the different multiplets. In this case, where the vev of the NGBs is zero, the mixing and mass corrections will affect only to those resonances which transform under SO(6) in the same representation. For example, those components of $Q\sim {\bf21}$ which transform under the representation $\bf15$ of SO(6) will mix with the components of $U\sim {\bf35}$ which transform under SO(6) in the same way.

The mixing between the two sites generates the appearance of partially composite states. When the pNGBs have a null vev, this mixing will provide partially composite massive fermions and partially composite massless chiral fermions. After setting to zero all the spurious degrees of freedom of the site-0, the remaining massless chiral fermions will be identified with the SM fermions. The rotation angle diagonalising the mixing is defined by:
\begin{equation}\label{eq-thetapsi}
\tan(\theta_\psi)=\frac{f_0 \lambda_\psi}{m_\Psi}\ ,
\end{equation}
and the degree of compositeness of the SM fermions is given by $\sin{\theta_\psi}$. The masses of the components of the resonances mixing with the elementary fermions is re-scaled by a factor $1/\cos\theta_\psi$, whereas the components that do not mix (usually called custodians) keep their original mass. For this reason the custodians are the lightest fermion resonances.

The SM fermions will get a mass only when the Higgs boson acquires a non-zero vev. This mass depends on the mixing with the composite resonances and on the Yukawa couplings of site-1, in agreement with the partial compositeness scenario. For a SM fermion $\psi$, the Yukawa coupling and the mass will be approximately given by:
\begin{align}
y_{\psi}\sim y_{\hat\Psi\bf r} \sin(\theta_{\psi})\sin(\theta_{\hat{\psi}})\ ,\qquad
m_{\psi}\sim y_\psi v,
\label{eq-PartComp}
\end{align}
where $v\cong246$ GeV is the EW scale, $\Psi$ is the resonance which contains the partner of $\psi_L$, $\hat{\Psi}$ contains the partner of $\psi_R$, $\bf r$ is the representation of SO(6) in common between $\Psi$ and $\hat{\Psi}$. In the present model we have omitted the label ${\bf r}$ since it takes only one value, ${\bf 15}$ for quarks and ${\bf 6}$ for leptons, as can be seen from Eqs.~(\ref{eq-ferm-rep}) and~(\ref{eq-dec-76}).

In the phenomenologically relevant situation where only the Higgs boson acquires a non-zero vev, the only preserved symmetry will be SU(3)$\times$U(1)$_{em}$. As a consequence, a much more complex pattern of mixing among fermions with equal electric charge is expected. In sec.~\ref{sec-ns} we will explain this pattern in more detail and the general mass matrices can be found in App.~\ref{ap-2site}.

\subsection{Low energy effective theory}
In order to study the low energy physics, it is useful to give an effective description below the scale of the resonances. This can be done by integrating the heavy states at tree level, keeping the elementary fields and the NGBs. 

As explained in the beginning of sec.~\ref{sec-2site}, to simplify our calculations we extend the symmetry of the elementary sector, site-0, to SO(7), by adding spurious degrees of freedom, that are non-dynamical and are set to zero in the end of the calculation. The elementary fermions are embedded in the representations of table~\ref{t-ferm}, and called $\psi_i$, as in Eq.~(\ref{eq-Lmix}).  

By making use of the CCWZ formalism of Refs.~\cite{Coleman:1969sm,Callan:1969sn}, it is possible to write an effective Lagrangian relying just on symmetry considerations. To quadratic order on the elementary fields:
\begin{align}\label{eq-leff}
{\cal L}_{\rm eff}\supset \
&\frac{f^2}{4}d^{\hat a}_\mu d^{\hat a,\mu}+
\sum_{{\bf r}={\bf 6},{\bf 15}}\Pi_{\bf r}(p^2)(U^\dagger a_\mu)_{\bf r}(U^\dagger a^\mu)_{\bf r} 
+\sum_{i=q,u,d,\ell,e}\sum_{\bf r}\Pi_{\bf r}^i(p^2)\overline{(U^\dagger\psi_i)}_{\bf r}\pslash
(U^\dagger\psi_i)_{\bf r}
\nonumber\\
&+\sum_{i=u,d}\sum_{\bf r}M_{\bf r}^i(p^2)\overline{(U^\dagger\psi_q)}_{\bf r}
(U^\dagger\psi_i)_{\bf r}
+\sum_{\bf r}M_{\bf r}^e(p^2)\overline{(U^\dagger\psi_\ell)}_{\bf r}
(U^\dagger\psi_e)_{\bf r} \ .
\end{align}
The subindex ${\bf r}$ takes values over the irreducible representations of SO(6) in a decomposition of an SO(7) representation ${\bf R}$: ${\bf R}\sim\oplus{\bf r}$, as explicitly shown in Eq.~(\ref{eq-dec-76}). The subindex in $(\phi)_{\bf r}$ has been used to denote the projection of the field $\phi$, transforming in a representation of SO(7), on representations of SO(6), whereas the product $(\phi)_{\bf r}(\phi')_{\bf r}$ is assumed to be projected on an SO(6)-singlet. The NGB-matrix $U$ is multiplied by an elementary field, thus it must be taken in the representation of that field. The form-factors $\Pi_{\bf r}(p^2)$ and $M_{\bf r}(p^2)$ codify the dynamics of the resonances that were integrated-out, they are independent of the NGBs. They can be computed integrating-out the resonances at site-1 and we show them explicitly in App.~\ref{ap-2site}. ${\cal L}_{\rm eff}$ also contains the usual kinetic terms of the elementary fields, that have not been written in Eq.~(\ref{eq-leff}).

By keeping just the dynamical elementary fields, the EW gauge and the quark sectors of Eq.~(\ref{eq-leff}) reduce to:
\begin{align}
{\cal L}_{\rm eff}\supset 
&\frac{1}{2}[Z_w+\Pi_w(p^2)]w_\mu^iw^{\mu i}+\frac{1}{2}[Z_b+\Pi_b(p^2)]b_\mu b^\mu+\Pi_{ib}(p^2)w_\mu^i b^\mu
\nonumber\\
&+\bar q_L\pslash(Z_q+\Pi_q)q_L+\sum_{\psi=u,d}[\bar\psi_R\pslash(Z_\psi+\Pi_\psi)\psi_R+\bar q_LM_{q\psi}\psi_R+{\rm h.c.}] \ .
\label{eq-leff1}
\end{align}
where $Z$ stands for non-canonical normalization of the kinetic terms: $Z_w=1/g_0^2$, $Z_b=1/{g_0'}^2$~\footnote{In the previous section we have set for the fermions $Z_\psi=1$.}. Taking $Z_\psi\to\infty$ one can study the limit of decoupling of the elementary sources, that will useful for the study of the one-loop potential. 
Although for simplicity we have shown only the quark sector, the lepton sector can be included straightforwardly. 

The boson form-factors, to all orders in the NGBs, are given by:
\begin{align}
&\Pi_w=\Pi_{6}\left(p^{2}\right)i_w^6+\Pi_{15}\left(p^{2}\right)i_w^{15}
\ ,\qquad
\Pi_b=\Pi_{6}\left(p^{2}\right)i_b^6+\Pi_{15}\left(p^{2}\right)i_b^{15}
\ ,\nonumber\\
&\Pi_{jb}=\left[\Pi_{6}\left(p^{2}\right)-\Pi_{15}\left(p^{2}\right)\right] i_{jb} \ ,
\label{eq-pijb}
\end{align}
where the NGB dependent invariants $i$ are:
\begin{align}
&i_w^6=\frac{H^{2}}{2\Gamma^{2}}\sin^{2}\left(\frac{\Gamma}{f}\right)\ ,&
&i_{1b}=\frac{3}{\Gamma^{2}}\sqrt{\frac{2}{17}}\sin^{2}\left(\frac{\Gamma}{f}\right)\left(h_{1}h_{3}-h_{2}h_{4}\right)
\ ,\nonumber\\
&i_w^{15}=\frac{1}{4\Gamma^{2}}\left(H^{2}\cos\left(2\frac{\Gamma}{f}\right)+4\Gamma^{2}-H^2\right)\ ,&
&i_{2b}=-i_{ib}
\ ,\nonumber\\
&i_b^6=\frac{16\Gamma^{2}-7H^{2}}{34\Gamma^{2}}\sin^{2}\left(\frac{\Gamma}{f}\right)\ ,&
&i_{3b}=\frac{3}{\sqrt{17}\Gamma^{2}}\sin^{2}\left(\frac{\Gamma}{f}\right)\left[H^{2}-2\left(h_{3}^{2}+h_{4}^{2}\right)\right]
\ ,\nonumber\\
&i_b^{15}=\frac{\Gamma^{2}\left(18+16\cos^2\left(\Gamma/f\right)\right)+7H^{2}\sin^2\left(\Gamma/f\right)}{34\Gamma^{2}}\ ,&
&
\end{align}
 and we use the definition $H^2=h_{1}^2+h_{2}^2+h_{3}^2+h_{4}^2$.

The fermion form-factors are:
\begin{align}
&\Pi_q=\left(\begin{array}{cc}\Pi_{uu}&\Pi_{ud}\\\Pi_{ud}^*&\Pi_{dd}\end{array}\right)\ ,&
&M_{q\psi}=\left(\begin{array}{c}M_{u\psi}\\M_{d\psi}\end{array}\right)\ ,
\end{align}
with
\begin{align}
&\Pi_{\psi\psi'}=\Pi^q_6(p^2) i_{q\psi\psi'}^6+\Pi^q_{15}(p^2) i_{q\psi\psi'}^{15}\ ,&\psi,\psi'=u,d\ ,&
\nonumber\\
&\Pi_{\psi}=\Pi^\psi_{10}(p^2) i_{\psi}^{10}+\Pi^\psi_{\overline{10}}(p^2) i_{\psi}^{\overline{10}}+\Pi^\psi_{15}(p^2) i_{\psi}^{15}\ ,& \psi=u,d\ ,&
\nonumber\\
&M_{\psi\psi'}=M_{15}^{\psi'}(p^2) i_{\psi\psi'}\ ,&\psi,\psi'=u,d\ .&
\end{align}
The NGB dependent invariants of $\Pi_q$ are:
\begin{align}
&i_{quu}^6= \frac{\Gamma^{2}-h_{1}^{2}-h_{2}^{2}}{2\Gamma^{2}}\sin^{2}\left(\frac{\Gamma}{f}\right)\ ,&
&i_{quu}^{15}= \frac{\Gamma^{2}\left(1+\cos^2\left(\frac{\Gamma}{f}\right)\right)+\left(h_{1}^{2}+h_{2}^{2}\right)\sin^2\left(\frac{\Gamma}{f}\right)}{2\Gamma^{2}}\ ,
\nonumber\\
&i_{qdd}^6=\frac{\Gamma^{2}-h_{3}^{2}-h_{4}^{2}}{2\Gamma^{2}}\sin^{2}\left(\frac{\Gamma}{f}\right)
\ ,&
&i_{qdd}^{15}=\frac{1}{4\Gamma^{2}}\left[\left(\Gamma^{2}-h_{3}^{2}-h_{4}^{2}\right)\cos\left(2\frac{\Gamma}{f}\right)+3\Gamma^{2}+h_{3}^{2}+h_{4}^{2}\right]\ ,
\nonumber\\
&i_{qud}^6=\frac{\left(h_{1}-i\, h_{2}\right)\left(h_{3}-i\, h_{4}\right)}{2\Gamma^{2}}\sin^{2}\left(\frac{\Gamma}{f}\right)
\ ,&
&i_{qud}^{15}=-i_{qud}^6 \ ,
\end{align}
for $\Pi_{\psi}$ we obtain:
\begin{align}
&i_{u}^{10}=i_{d}^{10}=\sin^{4}\left(\frac{\Gamma}{2f}\right)
\ ,&
&i_{u}^{\overline{10}}=i_{d}^{\overline{10}}=\cos^{4}\left(\frac{\Gamma}{2f}\right)
\ ,
&i_{u}^{15}=i_{d}^{15}=\frac{\sin^{2}\left(\frac{\Gamma}{f}\right)}{2}\ ,
\end{align}
and for $M_{\psi\psi'}$:
\begin{align}
&i_{uu}=\frac{h_{4}+i\, h_{3}}{2\Gamma}\sin\left(\frac{\Gamma}{f}\right)
\ ,&
&i_{ud}= \frac{h_{2}+i\, h_{1}}{\sqrt{2}\Gamma}\sin\left(\frac{\Gamma}{f}\right)\ ,
\nonumber\\
&i_{du}=\frac{i\, h_{1}-h_{2}}{2\Gamma}\sin\left(\frac{\Gamma}{f}\right)
\ ,&
&i_{dd}= \frac{h_{4}-i\, h_{3}}{\sqrt{2}\Gamma}\sin\left(\frac{\Gamma}{f}\right)\ .
\label{eq-idd}
\end{align}
Using Eqs.~(\ref{eq-pijb}-\ref{eq-idd}) in (\ref{eq-leff1}) one can obtain ${\cal L}_{\rm eff}$ to all order in the NGBs.

Evaluating the NGBs on their vev, Eq.~(\ref{eq-vev}), and denoting the corresponding form-factors as $\hat\Pi$, we obtain:
\begin{align}
{\cal L}_{\rm eff}\supset& 
\frac{1}{2}[Z_w+\hat\Pi_w(p^2)]w_\mu^iw^{\mu i}+\frac{1}{2}[Z_b+\hat\Pi_b(p^2)]b_\mu b^\mu+\hat\Pi_{3b}(p^2)w_\mu^3 b^\mu
\nonumber\\
&+\sum_{\psi=u,d,e}\left\{\bar\psi_L\hat M_{\psi_L\psi_R}(p^2)\psi_R+{\rm h.c.}+\sum_{X=L,R}\bar\psi_X\pslash [Z_{\psi_X}+\hat\Pi_{\psi_X}(p^2)]\psi_X\right\} \ .
\end{align}
The kinetic term of the NGBs give a contribution to the gauge form-factors that lead to mass terms for EW gauge bosons. From them one can obtain a matching condition of the SM Higgs-vev, that reads:
\begin{equation}\label{eq-matchvsm}
v^2=\xi f^2=(246\ {\rm GeV})^2\ , \qquad \xi=\sin(\langle h\rangle/f)^2 \ .
\end{equation}

The form-factors can be expressed in terms of the invariants evaluated on the Higgs vev and the form-factors of the vacuum $\Phi_0$. For the gauge fields we obtain: 
\begin{align}
&\hat\Pi_w=\Pi_{15}+\frac{\xi}{2}(\Pi_{6}-\Pi_{15})
\ ,&
&\hat\Pi_b=\Pi_{15}+\frac{\xi}{2}\frac{9}{17}(\Pi_{6}-\Pi_{15})
\ ,
\nonumber\\
&\hat\Pi_{3b}=-\frac{\xi}{2}\frac{3}{\sqrt{17}}(\Pi_{6}-\Pi_{15})
\ ,&&
\end{align}
and for the fermions:
\begin{align}
&\hat\Pi_{u_L}=\Pi^q_{15}+\frac{\xi}{2}(\Pi^q_{6}-\Pi^q_{15})
\ ,&
&\hat\Pi_{d_L}=\Pi^q_{15}
\ ,
\nonumber\\
&\hat\Pi_{u_R}=\frac{2-\xi-2\sqrt{1-\xi}}{4}\Pi^u_{10}+\frac{2-\xi+2\sqrt{1-\xi}}{4}\Pi^u_{\overline{10}}+\frac{\xi}{2}\Pi^u_{15}
\ ,&
&\hat\Pi_{d_R}=\hat\Pi_{u_R}(u\to d)
\ ,
\nonumber\\
&\hat M_{u_Lu_R}=\frac{\sqrt{\xi}}{2}M^u_{15}
\ ,&
&\hat M_{d_Ld_R}=\sqrt{\frac{\xi}{2}}M^d_{15}
\ .
\end{align}

The spectrum of up- and down-type states excited by the the elementary fermions, including the light states associated with the SM degrees of freedom as well as the resonances, can be obtained from:
\begin{equation}\label{eq-specf}
p^2[Z_{\psi_L}+\hat\Pi_{\psi_L}(p^2)][Z_{\psi_R}+\hat\Pi_{\psi_R}(p^2)]-|\hat M_{\psi_L\psi_R}(p^2)|^2=0\ .
\end{equation}
An approximate expression for the mass of the lightest fermion, the would-be zero mode, can be obtained by expanding Eq.~(\ref{eq-specf}) in powers of momentum to first order. Taking the derivative of the mass with respect to $\langle h\rangle$, one can obtain an approximate expression for the Yukawa coupling with $h$.~\cite{Carena:2014ria}  A very useful expression for the Higgs phenomenology is the ratio:
\begin{equation}\label{eq-yom}
\frac{y_\psi^{(0)}}{m_\psi^{(0)}}\simeq \frac{F_\psi(\xi)}{\sqrt{\xi}f}\left[1+{\cal O}\left(\xi\frac{\lambda_{\psi_L}^2f^2}{m_\Psi^2},\xi\frac{\lambda_{\psi_R}^2f^2}{m_\Psi^2}\right)\right] \ .
\end{equation}
In the present model we obtain the leading order correction:
\begin{equation}\label{eq-Fpsi}
F_u=F_d=F_e=\sqrt{1-\xi}\ ,
\end{equation}
where $F_e$ is for the charged leptons, and has been computed using the embeddings of Eq.~(\ref{eq-ferm-rep}). 

Eq.~(\ref{eq-yom}), together with Eq.~(\ref{eq-matchvsm}), must be compared with the SM result: $y^{\rm SM}_\psi/m_\psi=v^{-1}$. As has been discussed in several references, for example~\cite{Azatov:2011qy,Falkowski:2007hz,Carena:2014ria}, $F_u$ also give the dominant correction to the Higgs coupling to two gluons. 
In sec.~\ref{sec-hpheno} we show the next order in the expansion in powers of $\xi$ for the case of the top and bottom, that are important for the Higgs physics. Those corrections will allow us to understand some phenomenological results. 

Eq.~(\ref{eq-Fpsi}) gives the same result as in the MCHM with $(q,u,d)$ embedded in the representations $(5,10,10)$ and $(5,1,10)$, for that reason the Higgs phenomenology in the present model is similar to that case.

Following a similar procedure, it is possible to obtain an approximate expression for the masses of the $W$ and $Z$ bosons. Taking the first and second derivative of these approximate masses with respect to $\langle h\rangle$ leads to the couplings $VVh$ and $VVhh$. The result is the same as in the MCHM:
\begin{equation}\label{eq-gom}
\frac{g_{WWh}}{g m_W^{(0)}}\simeq \sqrt{1-\xi} \left[1+{\cal O}\left(\xi\frac{g_0^2}{g_1^2}\right)\right]\ ,
\qquad
\frac{g_{WWhh}}{g^2/2}\simeq 1-2\xi \left[1+{\cal O}\left(\xi\frac{g_0^2}{g_1^2}\right)\right] \ .
\end{equation}
In sec.~\ref{sec-hpheno} we will show the predictions for the next to leading terms in the 2-site model.


\section{Potential}\label{sec-potential}

The gauging of the EW group $SU\left(2\right)_{L}\times U\left(1\right)_Y$ in the site 0 and the mixing between elementary and composite fermions explicitly break the symmetry $SO\left(7\right)$ down to $SU\left(2\right)_{L}\times U\left(1\right)_Y$. This induces a radiative potential for the pseudo-Nambu-Goldstone bosons, which can be computed at one-loop level using the Coleman-Weinberg method\cite{Coleman:1973jx}.

To all order in the NGBs, the one-loop potential can be written as:
\begin{equation}\label{eq-vfull}
V=\int\frac{d^4p}{(2\pi)^4}
\left(-2N_{c}\ln\left[\frac{\det\left[\mathcal{A_{F}}\right]}{\det\left[\mathcal{A_{F}}\big|_0\right]}\right]+\frac{3}{2}\ln\left[\frac{\det\left[\mathcal{A_{B}}\right]}{\det\left[\mathcal{A_{B}}\big|_0\right]}\right]\right) \ ,
\end{equation}
where $N_c$ is the number of colors, the subindex 0 indicates that the NGBs must be evaluated to zero, and the operators $\mathcal{A_{F}}$ and $\mathcal{A_{B}}$ are given by:
\begin{align}
\mathcal{A_{F}}=&\left(\begin{array}{cccc}
\pslash(Z_q+\Pi_{uu}) & \pslash\,\Pi_{ud} & M_{uu} & M_{ud} \\
\pslash\,\Pi_{ud}^{*} & \pslash(Z_q+\Pi_{dd}) & M_{du} & M_{dd} \\
M_{uu}^{*} & M_{du}^{*} & \pslash(Z_u+\Pi_u) & 0 \\
M_{ud}^{*} & M_{dd}^{*} & 0 & \pslash(Z_d+\Pi_d)
\end{array}\right)
\ ,\nonumber\\
\mathcal{A_{B}}=&\left(\begin{array}{cccc}
Z_wp^2+\Pi_w & 0 & 0 & \Pi_{1b}\\
0 & Z_wp^2+\Pi_w & 0 & \Pi_{2b}\\
0 & 0 & Z_wp^2+\Pi_w & \Pi_{3b}\\
\Pi_{1b} & \Pi_{2b} & \Pi_{3b} & Z_b p^2+\Pi_b
\end{array}\right) \ .
\end{align}
The contributions of $\mathcal{A_{F}}\big|_0$ and $\mathcal{A_{B}}\big|_0$ are independent of the NGBs and subtract a constant divergent term. 

The form-factors are proportional to the fermion mixing squared: $\lambda_\psi\lambda_{\psi'}$. Therefore, since we assumed that only $q_L$ and $u_R$ of the third generation have large mixing, we will not consider the effect of the other fermions on the potential. For the analysis of the bottom quark phenomenology we will include the elementary $b_R$ and its composite partner, neglecting its impact on the potential.

By making use of the form-factors of the 2-site theory, it is straightforward to check the convergence of the one-loop potential for low and large Euclidean momentum. The finiteness of $V$ can be understood by the following argument~\cite{Carena:2014ria}: a non-vanishing potential requires insertions of the Yukawa $y_\Psi$ and of the mixing $\lambda_\psi$, but it also requires insertions of the composite masses $m_\Psi$, since a chiral flip is needed, leading to $V\sim(f_0\lambda_\psi f_1 y_\Psi m_\Psi)^2$. Thus, by simple power counting, the one-loop potential is finite. If one includes operators like $f_1 y'_{U}[(\bar Q_R U_1)_{\bf 15}(U_1^\dagger U_L)_{\bf 15}]_{\bf1}$, that have the opposite chirality structure compared with those of Eq.~(\ref{eq-L1}), the one-loop potential becomes logarithmically divergent, since $V\sim(f_0\lambda_\psi f_1 y'_U)^2$. For that reason we have excluded those operators, other possibility is to consider a theory with three or more sites~\cite{Panico:2011pw}, or an extra dimensional theory~\cite{Contino:2003ve}.

\subsection{EW symmetry breaking}
To study the conditions for EWSB we find it useful to consider some limits of the potential. The first of them is the expansion in powers of the $H$ and $\chi$ fields up to fourth order. Using the invariants computed in the previous section, the expantion of the potential acquires the following shape: 
\begin{equation}
V= m_H^2 H^{2}+m_\chi^{2}\, \chi^{2}+\lambda_H\, H^{4}+\lambda_{H\chi}H^{2}\, \chi^{2}+\lambda_\chi\chi^{4}+\mathcal{O}(\phi^6),
\label{E:potO4}
\end{equation}
which is the most general expression allowed by the remnant symmetry $SU(2)_L\times U(1)_Y$. The ${\cal O}(\phi^6)$ stands for higher order terms, made from the product of both scalars, that are obtained after the expansion of the trigonometric functions of the invariants. 

The size of the quadratic and quartic coefficients can be estimated by doing an expansion in powers of the mixing, up to accidental cancellations the result is~\cite{Panico:2012uw,DaRold:2018moy}:
\begin{equation}\label{eq-estimateV}
m_\phi^2\sim \frac{\lambda_\psi^2 m_1^2}{16\pi^2}\ , 
\qquad
\lambda\sim \frac{\lambda_\psi^2 m_1^2}{16\pi^2f^2}\ , 
\end{equation}
where we are assuming that the sector of resonances, site-1, can be characterised by one scale $m_1$ and one coupling $g_1$, with $f\sim m_1/g_1$, and $\phi$ stands for both scalars. More complicated situations have been analysed, for example, in Ref.~\cite{Panico:2012uw}.

All the coefficients of the Eq.~(\ref{E:potO4}) can be expressed as integrals in momentum space of the form-factors of the effective theory defined in the previous section. The quadratic coefficients are simple enough to be studied by inspection:
\begin{align}
m_H^{2}=&\int\Bigg(N_{c}\left(\frac{\left|M_{u}\right|^{2}}{2\left(\Pi_{\overline{10}}^{u}+Z_{u_{R}}\right)\left(\Pi_{15}^{q}+Z_{q_{L}}\right)}-\frac{\left(\Pi_{6}^{q}-\Pi_{15}^{q}\right)}{\left(\Pi_{15}^{q}+Z_{q_{L}}\right)}-\frac{\left(\Pi_{15}^{u}-\Pi_{\overline{10}}^{u}\right)}{\left(\Pi_{\overline{10}}^{u}+Z_{u_{R}}\right)}\right)\nonumber\\
&+\frac{9}{34}\left(\Pi_{6}-\Pi_{15
}\right)\frac{\left(10\, \Pi_{15}+17 p^2 Z_{B}+3 p^2 Z_{W}\right)}{\left(\Pi_{15}+2 p^2 Z_{B}\right)\left(\Pi_{15}+2 p^2 Z_{W}\right)}\Bigg)\frac{d^{4}p}{\left(2\pi\right)^{4}}\nonumber\\
m_\chi^{2}=&\int\Bigg(-N_{c}\left(2\frac{\left(\Pi_{6}^{q}-\Pi_{15}^{q}\right)}{\left(\Pi_{15}^{q}+Z_{q_{L}}\right)}+\frac{\left(\Pi_{15}^{u}-\Pi_{\overline{10}}^{u}\right)}{\left(\Pi_{\overline{10}}^{u}+Z_{u_{R}}\right)}\right)+\frac{12}{17}\frac{\left(\Pi_{6}-\Pi_{15}\right)}{\left(\Pi_{15}+2 p^2 Z_{B}\right)}\Bigg)\frac{d^{4}p}{\left(2\pi\right)^{4}}.
\label{eq-intm2}
\end{align} 
Both coefficients receive contributions from the boson and fermion sectors of the theory. Also, in most Composite Higgs Models the relation $\Pi_{6}>\Pi_{15}$ is verified, therefore the bosonic contributions tends to prevent the existence of non-trivial minima. Similar analysis can be done for the quartic terms.

The minima of the potential of Eq.~(\ref{E:potO4}) can be obtained straightforwardly. There can be a trivial minimum with $\langle h\rangle=0$, $\langle \chi\rangle=0$, as well as a non-trivial minimum with:
\begin{equation}
\langle h\rangle^2 = \frac{m_\chi^{2}\lambda_{H\chi}-2m_H^{2}\lambda_{\chi}}{4\lambda_H\lambda_\chi-\lambda_{H\chi}^{2}},\qquad
\langle \chi\rangle ^{2} = \frac{m_H^{2}\lambda_{H\chi}-2m_\chi^{2}\lambda_H}{4\lambda_H\lambda_\chi-\lambda_{H\chi}^{2}}.
\end{equation}
For phenomenological reasons, that is: to preserve an unbroken U(1)$_{em}$, we are interested in the situation where only $H$ has a vev, this depends on the region of the parameter space, and does not require extra tuning. 
Notice that, as usual in composite Higgs models, using the estimates of Eq.~(\ref{eq-estimateV}) one obtains that, for a separation between $v$ and $f$, a tuning at least of order $1/\xi$ is required~\cite{Panico:2012uw}.~\footnote{See also Ref.~\cite{Csaki:2018zzf} for recent analysis of the tuning in Composite Higgs Models.} 

For $\langle \chi\rangle=0$, the masses of the scalar fields can be estimated by using Eqs.~(\ref{E:potO4}) and~(\ref{eq-estimateV}). The mass of  the exotic scalar is given by the estimate of Eq.~(\ref{eq-estimateV}), up to corrections of order ${\cal O}(\xi)$, whereas for the Higgs there is an extra suppression: $m_h^2\sim \xi \lambda_\psi^2 m_1^2/(16\pi^2)$.
Thus, for the selected vacuum, $m_\chi$ results higher than $m_h$.

Since the NGB matrix can be expressed in terms of $\sin(h/f)$, it is also interesting to consider an expansion of the Higgs potential in powers of $\sin(h/f)$. For the Higgs vev, in the limit of $\xi\ll 1$:
\begin{equation}\label{eq-vsin4}
V\simeq a \sin\left(\frac{\langle h\rangle}{f}\right)^2+b \sin\left(\frac{\langle h\rangle}{f}\right)^4 \ ,
\end{equation}
where $a$ and $b$ can be expressed in terms of integrals of the correlators, similar to Eq.~(\ref{eq-intm2}). This equation will be useful to obtain approximate expressions for the Higgs self-couplings, as we will show in sec.~\ref{sec-hpheno}.

\subsection{Numerical results}
In order to explore the parameter space, we computed numerically the potential for different parameter values. The gauge couplings $g_{0}$ and $g_{0}'$ were fixed as functions of $g_{1}$ in order to obtain the SM couplings at EW scale: $g=0.65$ and $g'=0.35$. Therefore, the potential depends on 8 parameters: $f_0$, $f_1$, $m_U$, $m_Q$, $\theta_u$, $\theta_q$, $y_{U}$ and $g_1$. The first four have dimension of mass and the rest are dimensionless.

First, a random scan of the parameter space was performed. The explored ranges for the different parameters were $f_{0,1}\sim 1$ TeV, $m_{U,Q}\in (0.5,10)\text{ TeV}$, $\theta_{q,u}\in(0.4,\pi/2)$, $y_{U}\in(0.1,3)$ and $g_{1}\in(1,6)$. Only those points where the non trivial minimum was located at $\langle\chi\rangle=0$ and $0<\xi<1$ were analysed.

\begin{figure}[h!]
	\begin{center}
\includegraphics[width=0.49\textwidth]{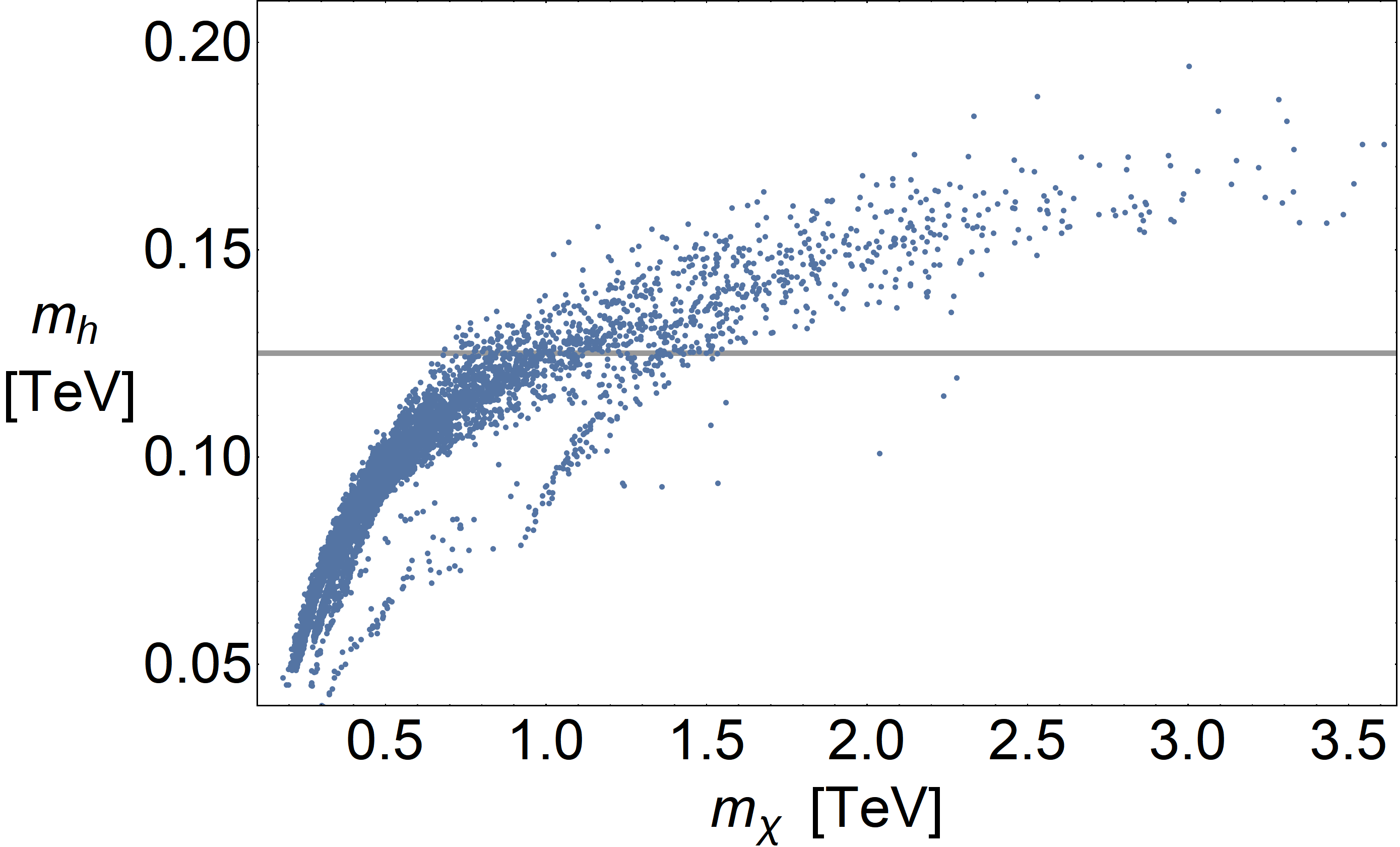}
\hskip0.2cm
\includegraphics[width=0.49\textwidth]{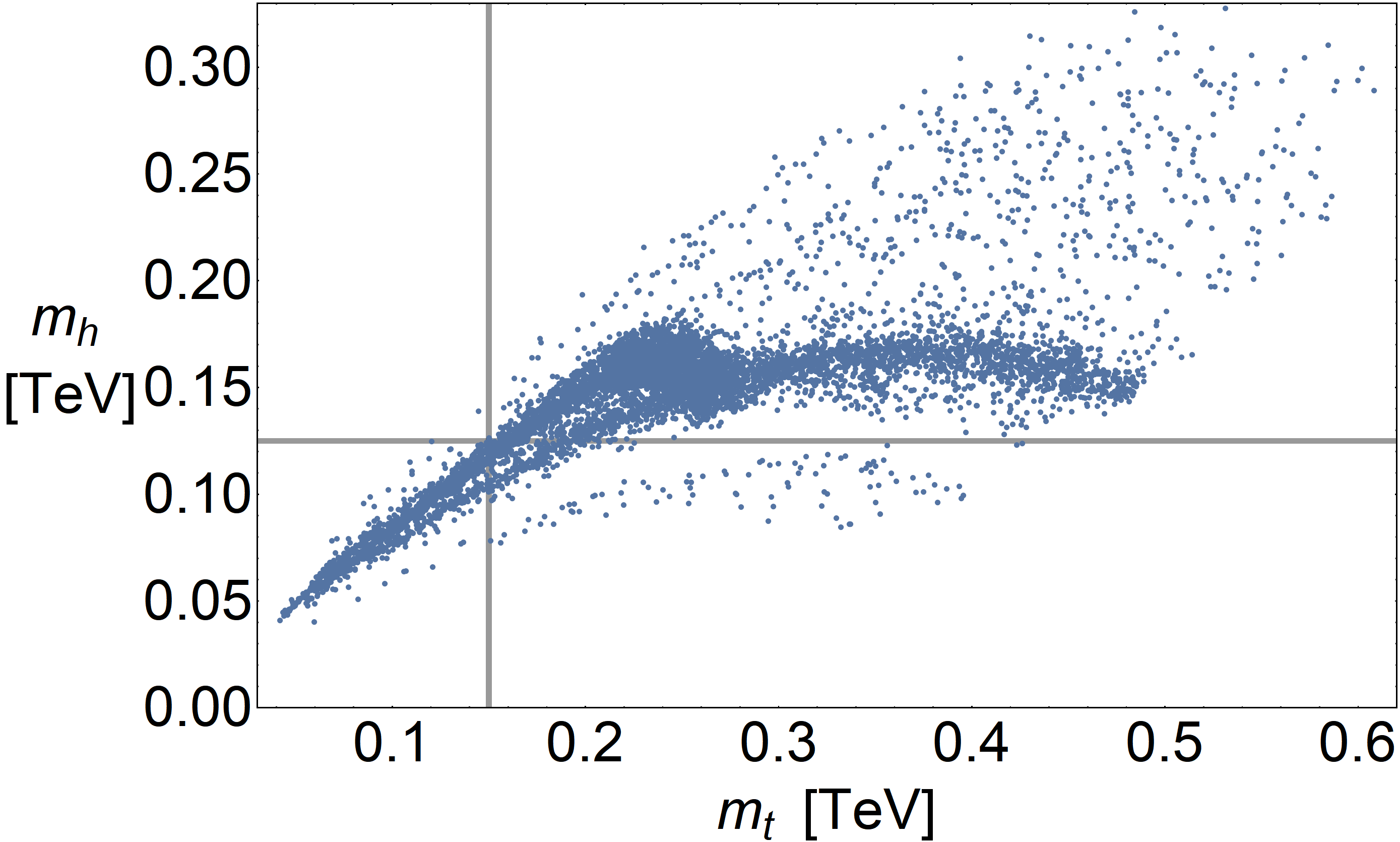}
\caption{Left panel: $h$ and $\chi$ masses, with a random scan of parameters as explained in the text, re-scaling the dimensionful quantities to fix $v=246$~GeV. Right panel: Higgs and top quark masses, with a random scan, fixing $f_0=1$~TeV. The gray lines correspond to $m_h=125$~GeV and $m_t=150$~GeV.}
\label{F:spectrum}
\end{center}
\end{figure}

We show our results in Fig.~\ref{F:spectrum}. The analysis of the points selected with the aforementioned criteria allowed us to extract the following conclusions: first, after re-scaling the dimensionful quantities such that $v= 246$~GeV, the Higgs mass acquires values between $50$~GeV and $200$~GeV, while the $\chi$ boson mass ranges from $100$~GeV to $4$~TeV. In particular, for those points where the Higgs mass is near $125$~GeV, the $\chi$ boson mass is around $1$~TeV. Second, if the dimensionful quantities are re-scaled such that $f_{0}=1$~TeV, the Higgs boson mass ranges in a similar interval than before and, for those points where it is near $125$~GeV, the top quark mass ranges between $150$~GeV and $200$~GeV.  

After re-scaling the set of randomly obtained points such that $v=246$ GeV, we selected a subset which had phenomenological interest. The criteria to select those points were: $f_{0}g_{1}>2$ TeV, $100\text{ GeV}<m_H<145$ GeV, $140\text{ GeV}<m_t<175$ GeV\footnote{The relevant masses should be the running masses at the scale where the resonances are integrated out. These would then be run down to the weak scale to match the experimental measurements. In order to simplify the scan, we have just chosen some wide ranges of acceptable values.} and $\xi<0.25$. 240 points were found to meet these requirements. This set of benchmark points (BP) were used to carry out phenomenological studies which will be showed in the following section.

We chose a specific point of the set of BP which fulfilled all the requirements exceptionally well and decided to explore systematically the parameter space around it. The point is defined by: $f_0=1.47$~TeV, $f_1=2.34$~TeV, $m_U=2.44$~TeV, $m_Q=1.26$~TeV, $\theta_u=0.79$, $\theta_q=1.37$, $y_{U}=2.52$ and $g_1=1.95$. In each point we computed the vev of the Higgs boson, its mass and the masses of the $\chi$ boson and the top quark. Below we show our results in regions where only two parameters were varied at the same time, while the other parameters were kept fixed. 

\begin{figure}[h!]
	\begin{center}
		\includegraphics[width=0.7\textwidth]{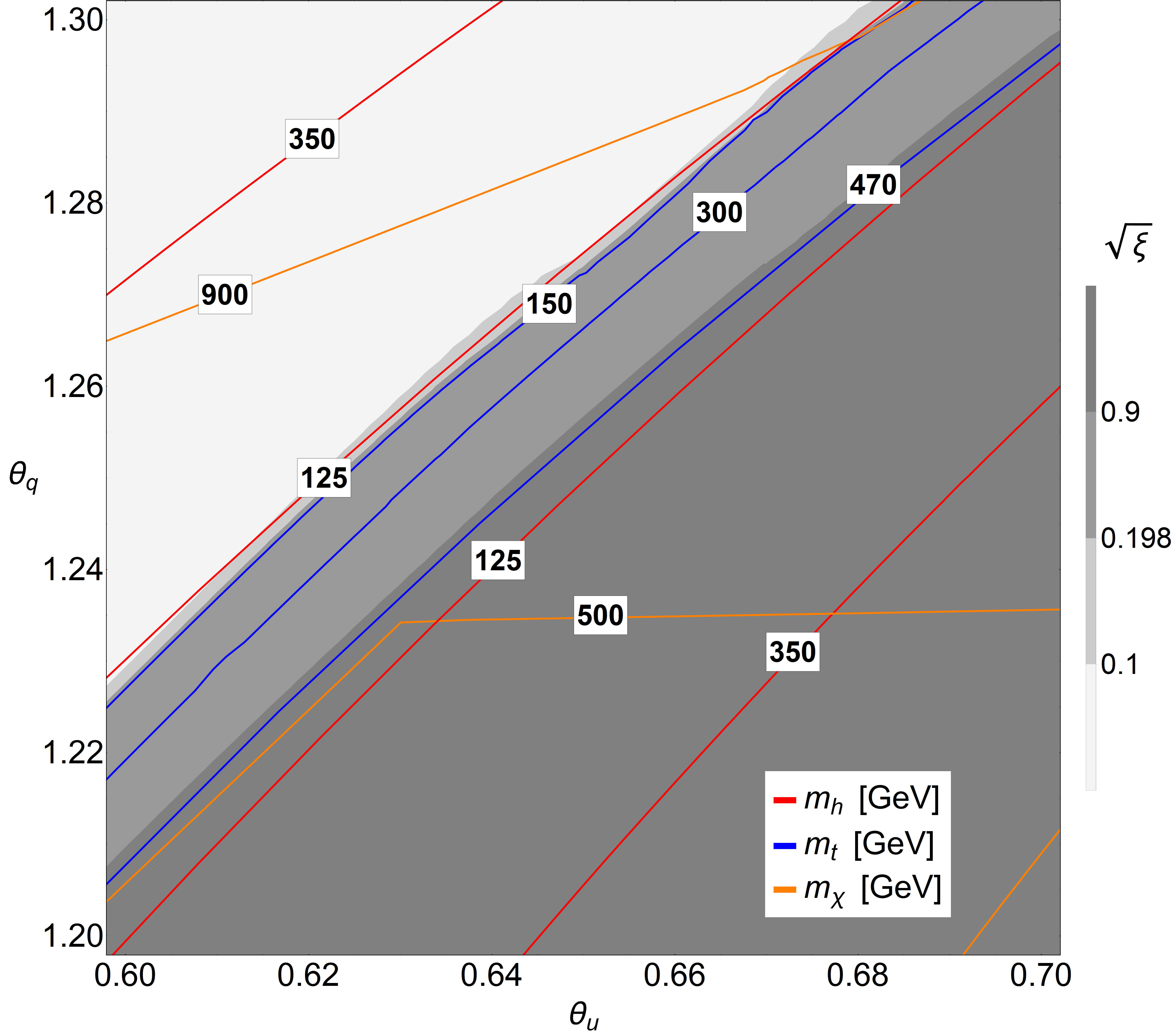}
		\caption{Contour plots of $\sqrt{\xi}$ and the masses of the Higgs boson ($m_H$), the top quark ($m_t$) and the $\chi$ boson ($m_\chi$) as functions of the mixing angles $\theta_{u}$ and $\theta_{q}$. The showed region is just a small portion of the explored parameter space where phenomenologically interesting points can be found. The value $\sqrt{\xi}=0.198$ corresponds to $v=246$ GeV.}
		\label{F:cnIntAngUQdetail}
	\end{center}
\end{figure}

In Fig.~\ref{F:cnIntAngUQdetail} we show the contour plots of some phenomenologically relevant quantities as functions of the mixing angles in a small portion of the explored region. There is a zone where all the quantities take values near the measured ones. This means that the Higgs mass is near $125$ GeV, the top quark mass is near $150$ GeV, the $\chi$ boson mass is high enough to fulfill the experimental constraints~\cite{Chala:2018opy} and the EW scale $v$ is near $246$ GeV\footnote{The EW scale $v$ is closely related to the EW boson masses, like in the SM. Ensuring that $v$ is close to $246$ GeV means that those masses are near their measured values.}. The shape of that phenomenologically interesting region depends on the exact values of all the other parameters.

\begin{figure}[h!]
	\begin{center}	\includegraphics[width=0.7\textwidth]{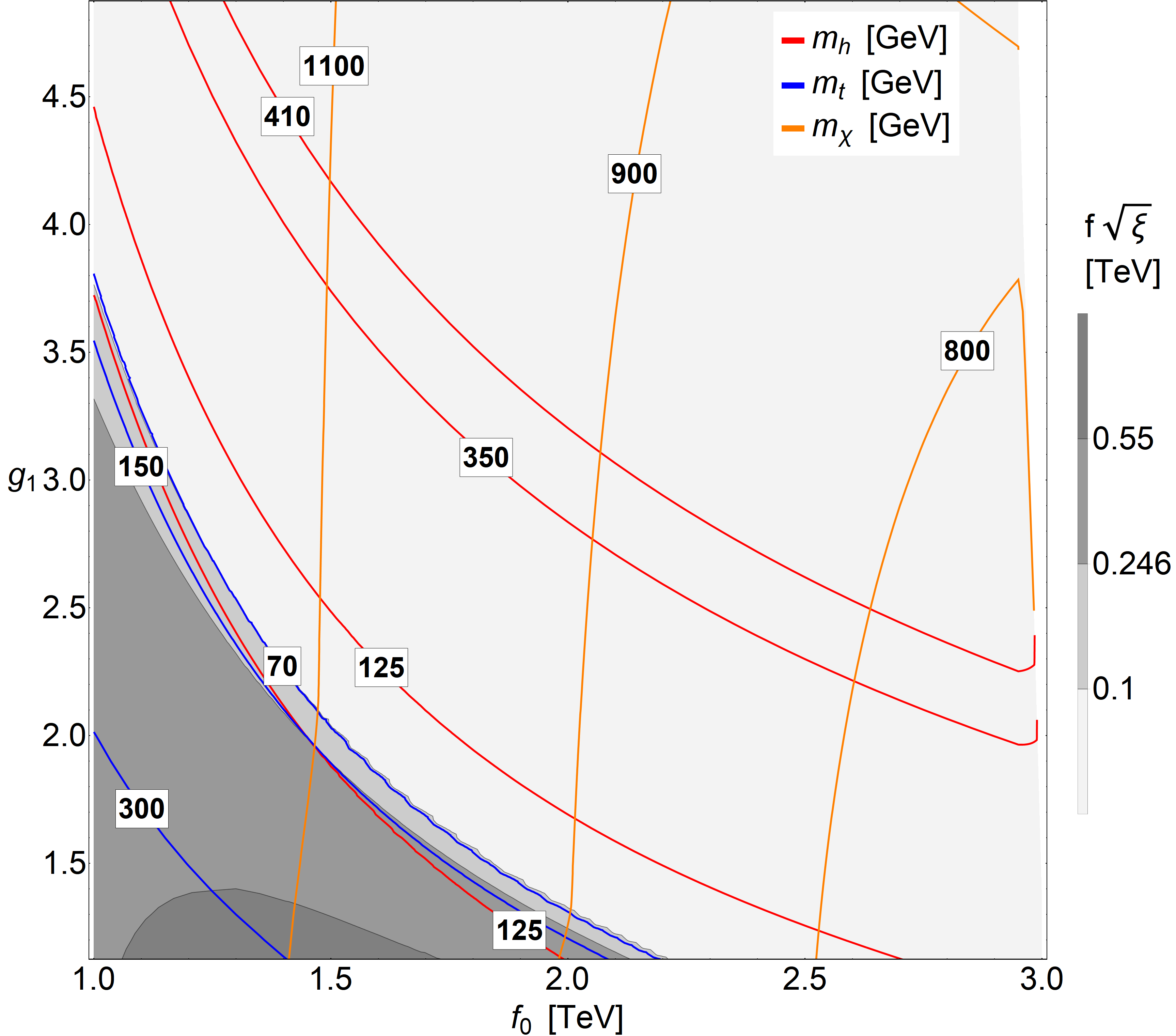}
		\caption{Contour plots of $f \sqrt{\xi}=v$ and the masses of the Higgs boson ($m_H$), the top quark ($m_t$) and the $\chi$ boson ($m_\chi$) as functions of $f_0$ and $g_1$. The value $f\sqrt{\xi}=0.246$ TeV is the phenomenologically interesting value.}
		\label{F:cvIntf0g1}
	\end{center}
\end{figure}

If the scan is performed in the parameters $f_0$ and $g_1$, another phenomenologically interesting region is found. This can be seen in the Fig.~\ref{F:cvIntf0g1}, where all the quantities are near their measured values for $f_0\sim 1.5$ TeV and $g_1\sim 2$. Regions with similar phenomenological relevance are found in all the other scans, but their shapes and extensions depend on the values of the parameters which are fixed and on the parameters that are being scanned.

\section{Phenomenology}\label{sec-pheno}
In this section we describe the properties of the new states, their masses and quantum numbers. We also discuss the effects associated to compositeness in Higgs physics, as corrections to couplings, as well as production at LHC, total width and branching ratios. Finally we discuss briefly the stability of the resonnaces, the phenomenology of the pNGB $\chi$ and the main interactions for its production at LHC.

\subsection{New states}\label{sec-ns}
There are 21 real resonances of spin one. States with electric charge $0$, $\pm\frac{1}{3}$, $\pm\frac{2}{3}$, $\pm1$ and $\pm\frac{5}{3}$ are found among them. Non-integer charges appear because these resonances transform under $SO\left(7\right)$ in the same irreducible representation as the Left-handed doublet of quarks. Therefore, vector bosons with non-integer electric charge are unavoidable and become a distinctive signature of this model.

After EWSB, seven different values of masses can be identified in the spectrum of vector resonances. The Table \ref{T:especBosonVec} summarises the mentioned spectrum. There we have used the name $m_A^{i}$ for all the bosons with mass $m_i$. Since the vacuum must preserve the electromagnetic gauge symmetry, only the masses of some vector bosons with charge $0$ or $\pm 1$, which are allowed to mix with the elementary electroweak gauge bosons, can be corrected by mixing. The value of those corrections can not be computed analytically, but we have checked numerically that they are smaller than $0.1\%$ in the BP.

\begin{table}
\renewcommand*{\arraystretch}{1.35}
\begin{centering}
\begin{tabular}{|c|c|c|c|}
\hline 
Name & Mass & $|Q_{em}|$ & Multiplicity \tabularnewline[5pt]
\hline\rule{0mm}{5mm}
$m_A^{1}$ & $\frac{f_{0}g_{1}}{\sqrt{2}}$ & $\{0,1/3,2/3,1,5/3\}$ & $\{1,2,4,2,2\}$\tabularnewline[5pt]
\hline\rule{0mm}{5mm} 
$m_A^{{2}}$ & $f_{0}\sqrt{\frac{{g_{0}'}^{2}+g_{1}^{2}}{2}}+\epsilon$ & $0$ & $1$\tabularnewline[5pt]
\hline\rule{0mm}{5mm} 
$m_A^{{3}}$ & $f_{0}\sqrt{\frac{g_{0}^{2}+g_{1}^{2}}{2}}+\eta$ & $1$ & $2$\tabularnewline[5pt]
\hline\rule{0mm}{5mm} 
$m_A^{{4}}$ & $f_{0}\sqrt{\frac{g_{0}^{2}+g_{1}^{2}}{2}}+\Delta$ & $0$ & $1$\tabularnewline[5pt]
\hline\rule{0mm}{5mm} 
$m_A^{{5}}$ & $g_{1}\sqrt{\frac{f_{0}^{2}+f_{1}^{2}}{2}}$ & $\{2/3,0\}$ & $\{2,1\}$\tabularnewline[5pt]
\hline\rule{0mm}{5mm} 
$m_A^{{6}}$ & $g_{1}\sqrt{\frac{f_{0}^{2}+f_{1}^{2}}{2}}+\delta$ & $1$ & $2$\tabularnewline[5pt]
\hline\rule{0mm}{5mm} 
$m_A^{{7}}$ & $g_{1}\sqrt{\frac{f_{0}^{2}+f_{1}^{2}}{2}}+\alpha$ & $0$ & $1$\tabularnewline[5pt]
\hline 
\end{tabular}
\par
\end{centering}
\protect\caption{Spectrum of vector boson resonances with their masses and electric charges. $m_A^{i}$ is a generic denomination for all the vector resonances with the same mass. In cases with several charges, the multiplicities follow the order of the charges. $\epsilon$, $\eta$, $\Delta$, $\delta$ and $\alpha$ stand for corrections to the masses due to EWSB that can not be computed analytically (although it is possible to obtain simple expressions expanding in powers of $\xi$).}
\label{T:especBosonVec}
\end{table}

\begin{figure}[h!]
\begin{center}
\includegraphics[width=0.49\textwidth]{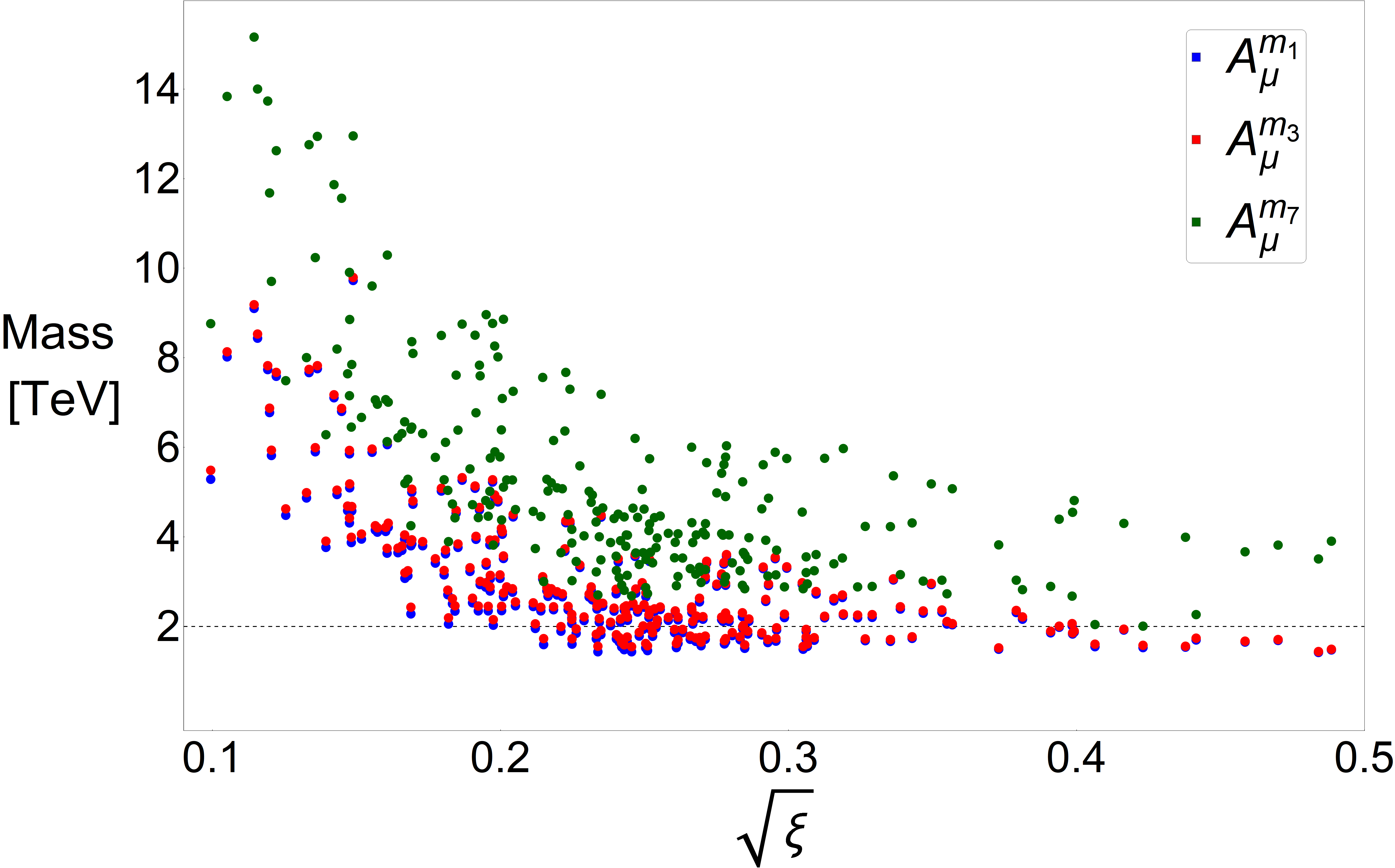}
\hskip0.1cm
\includegraphics[width=0.49\textwidth]{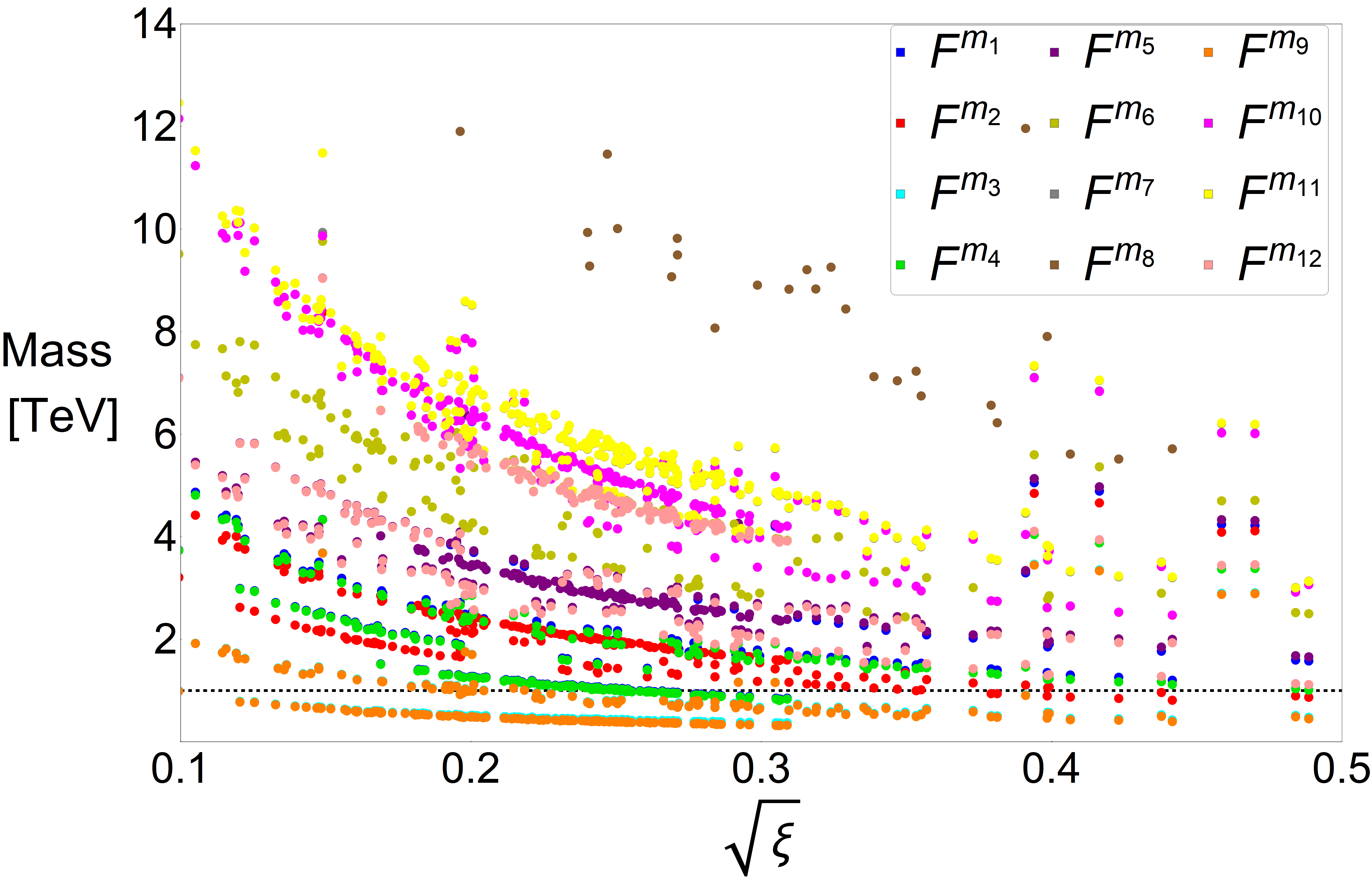}
\caption{On the left we show the masses of the resonances $m_A^{{1}}$, $m_A^{{3}}$ and $m_A^{{7}}$ in the BP as a function of $\sqrt{\xi}$. The notation agrees with Table \ref{T:especBosonVec}, the horizontal dotted line is at $m=2$~TeV. On the right we show the masses of the fermion resonances in the BP. The notation is the same than in Table \ref{T:especResFerm}. The horizontal dotted line is at $1$~TeV, many points of $m_{F}^{3}$ are over and slightly below $m_F^{{9}}$.}
\label{F:masResVecPI}
\end{center}
\end{figure}

In the Fig.~\ref{F:masResVecPI}, we show three of the vector resonance masses for the BP, where the EW scale is fixed in $v=246$ GeV. The non plotted masses would be hidden in the plot due to the smallness of the mixing corrections. The masses rise as $\xi$ decreases because the latter implies a bigger separation between the electroweak scale and the scale of the composite sector. The dotted horizontal line is located at $2$ TeV and represents in an illustrative way the current experimental bounds~\cite{Sirunyan:2018fki,Sirunyan:2018rfo}. As it can be seen in the Fig.~\ref{F:masResVecPI}, all the points with $\xi<0.04$ are above 2~TeV, whereas for $\xi>0.16$ there are always resonances with masses below 2~TeV

\begin{table}
\renewcommand*{\arraystretch}{1.35}
\begin{centering}
\begin{tabular}{|c|c|c|}
\hline 
Name & $Q_{em}$ &\begin{tabular}{@{}c@{}}Number of Dirac fermions\end{tabular} \tabularnewline
\hline 
$m_F^{{1}}$ & $\{0,\pm 1,-2/3\}$ & $\{2,2,1\}$\tabularnewline
\hline 
$m_F^{{2}}$ & $\{0,\pm 1/3,2/3,-2/3,\pm 1,\pm 5/3\}$ & $\{4,4,1,2,4,4\}$\tabularnewline
\hline 
$m_F^{{3}}$ & $2/3$ & $1$\tabularnewline
\hline 
$m_F^{{4}}$ & $2/3$ & $1$\tabularnewline
\hline 
\multirow{1}{*}{$m_F^{{5}}$} & $2/3$ & $1$\tabularnewline
\hline 
$m_F^{{6}}$ & $2/3$ & $1$\tabularnewline
\hline 
$m_F^{{7}}$ & $2/3$ & $1$\tabularnewline
\hline 
$m_F^{{8}}$ & $2/3$ & $1$\tabularnewline
\hline 
$m_F^{{9}}$ & $\{0,1/3,-2/3,\pm 1,\pm 5/3\}$ & $\{3,1,2,4,2\}$\tabularnewline
\hline 
$m_F^{{10}}$ & $\{0,1/3,-2/3,\pm 1,\pm 5/3\}$ & $\{3,1,2,4,2\}$\tabularnewline
\hline 
$m_F^{{11}}$ & $-\frac{1}{3}$ & $1$\tabularnewline
\hline 
$m_F^{{12}}$ & $-\frac{1}{3}$ & $1$\tabularnewline
\hline 
\end{tabular}
\par\end{centering}
\protect\caption{Spectrum of fermion resonances, partners of $q_L$ and $u_R$ of the third generation. $m_F^{{i}}$ is a generic name for all the fermions with the same mass. Where the electric charge is denoted with both signs, the amount of fermions in the third column is the sum of fermions with the positive charge and fermions with the negative charge. In those cases, the amount of fermions with a given sign in its electric charge is half the amount indicated.}
\label{T:especResFerm}
\end{table}

Taking into account the composite partners of $q_L$ and $u_R$ of the third generation, we have included 56 fermion resonances in our model. All of them are Dirac fermions which transform under $SU\left(3\right)_c$ in the fundamental representation. After EWSB and before elementary-composite mixing, 12 different masses are found in the fermion spectrum. 

In the Table \ref{T:especResFerm} we specify the amount of fermions with every mass and electric charge. As we did with the vector resonances, we use the symbol $m_F^{{i}}$ to denote  all the fermions with the same mass. The masses of $m^{1}_F$, $m^2_F$, $m^{9}_F$ and $m^{10}_F$ do not depend on the vev of the Higgs boson, neither on the mixing angles with the elementary fermions. 

Let us remind that these masses have been computed without including the Right-handed field for the bottom quark. In this case, the fermions $m_F^{{11}}$ and $m_F^{{12}}$ are states arising from the mixing among composite resonances and the elementary field $b_L$. Their masses can be computed analytically and they do not depend on the vev of the Higgs boson. If the elementary field $b_R$ were added, the masses of $m^{11}_F$ and $m^{12}_F$ would be corrected by the mixing angle $\theta_{d}$ and they would depend on the vev of the Higgs boson. Due to the smallness of the mixing angle $\theta_{d}$, required to reproduce the hierarchy between the top and bottom masses, we expect those corrections to be small.

The masses of the fermion resonances for the BP are shown in Fig.~\ref{F:masResVecPI}. The horizontal dotted line at $1$ TeV illustrates the current experimental bounds~\cite{Aaboud:2018uek,Aaboud:2018xuw}, which depend on details that we have not analysed in this work and may rise up to 2 TeV~\cite{DeLuca:2018mzn,ATLAS:2018yey,CMS:2016ybj}. As it was seen in the case of vector resonances, the masses grow as $\xi$ goes down, and only for $\sqrt{\xi}<0.2$ there are points of the parameter space where the experimental bound is satisfied.

The lightest fermion resonance in all the BP is $m_{F}^{9}$, which is a purely composite resonance and a custodian. This confirms the expectation discussed below Eq.~(\ref{eq-thetapsi}). The second lightest resonance is $m_{F}^{3}$, whose mass is corrected by mixing with the elementary top. This correction is small enough to make it hardly visible in the plot.

\subsection{Higgs phenomenology}\label{sec-hpheno}
In Fig.~\ref{F:spectrum} we showed the predictions for the Higgs boson mass. It is also possible to explore numerically the correlation between $m_h$ and $\xi$. We found that $m_h$ can be of order $125$~GeV for $\xi$ preferentially in the range $0.1^2-0.4^2$. There is no mixing between the Higgs and $\chi$ in any phenomenologically interesting scenario due to the preservation of the U(1)$_{em}$ gauge symmetry and the different electrical charges of the bosons.

We analysed the Yukawa coupling between the Higgs boson and the quarks and leptons. In order to study the coupling with the light fermions, we have added the corresponding resonances in the SO(7) representations of Eq.~(\ref{eq-ferm-rep}). We assumed that they do not modify the potential noticeably.\footnote{In App.~\ref{ap-2site} we show the correlators after inclusion of $b_R$ and its composite partner, it is also straightforward to compute the modifications of the mass matrices in this case.} The mass of the new resonances and their composite Yukawa couplings were assumed to be of the same order as those associated to the other fermion resonances. 
The mixing angle of the elementary right handed bottom, $\theta_{d}$, was tuned to match its mass to the measured value. For all the BP, that mixing angle turned out to be of order $0.01$. The mixing angles for the lighter fermions will be even smaller, according to Eq.\ref{eq-PartComp}.

Since the top and bottom quarks play a dominant role in Higgs physics, we discuss their couplings below. 
The leading order approximations of the Yukawa couplings, expanding in powers of $\xi$, were given in Eqs.~(\ref{eq-yom}) and (\ref{eq-Fpsi}). Below we show the next to leading order corrections to the Yukawa couplings of the quarks, expanding in powers of the mixing:
\begin{align}
\frac{y_u}{m_u}\simeq&\frac{F_u}{\sqrt{\xi}f}\left[1+\xi\frac{f_{1}^{2}\,y_{U}^{2}}{4}\left(\frac{\sin^{2}\left(\theta_{q}\right)}{m_U^{2}}-\frac{\sin^{2}\left(\theta_{u}\right)}{m_Q^{2}}\right)+\mathcal{O}\left(\sin^{4}\left(\theta_{q,u}\right)\right)\right]\ ,
\nonumber \\
\frac{y_d}{m_d}\simeq&\frac{F_d}{\sqrt{\xi}f}\left[1-\xi\frac{f_{1}^{2}\,y_{D}^{2}}{4}\frac{\sin^{2}\left(\theta_{d}\right)}{m_Q^{2}}+\mathcal{O}\left(\sin^{4}\left(\theta_{q,d}\right)\right)\right]\ .
\label{eq-yom2}
\end{align}

\begin{figure}[h!]
\begin{minipage}[b]{0.49\textwidth}
\includegraphics[width=\textwidth]{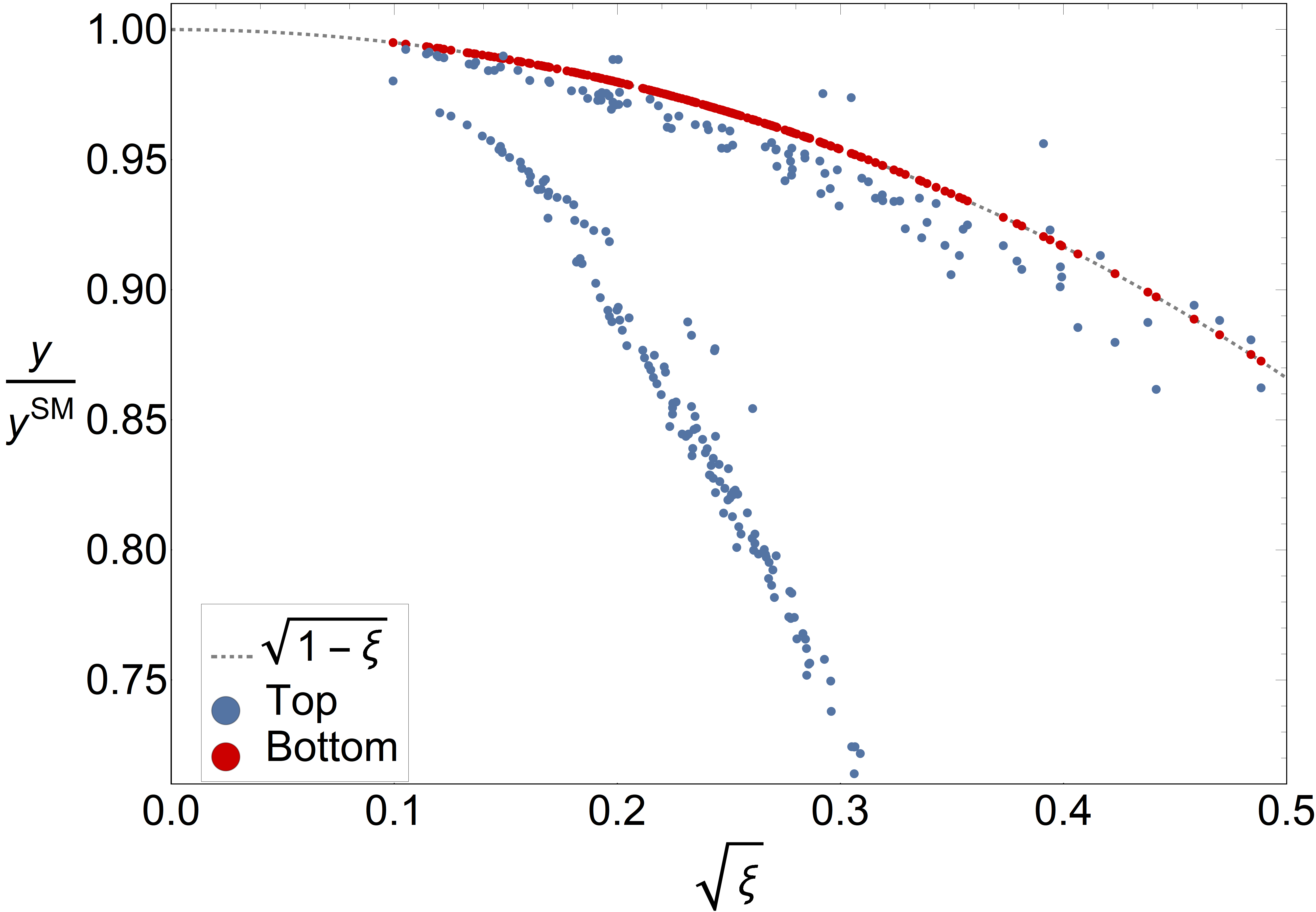}
\end{minipage}\hfill
\begin{minipage}[b]{0.49\textwidth}
\includegraphics[width=\textwidth]{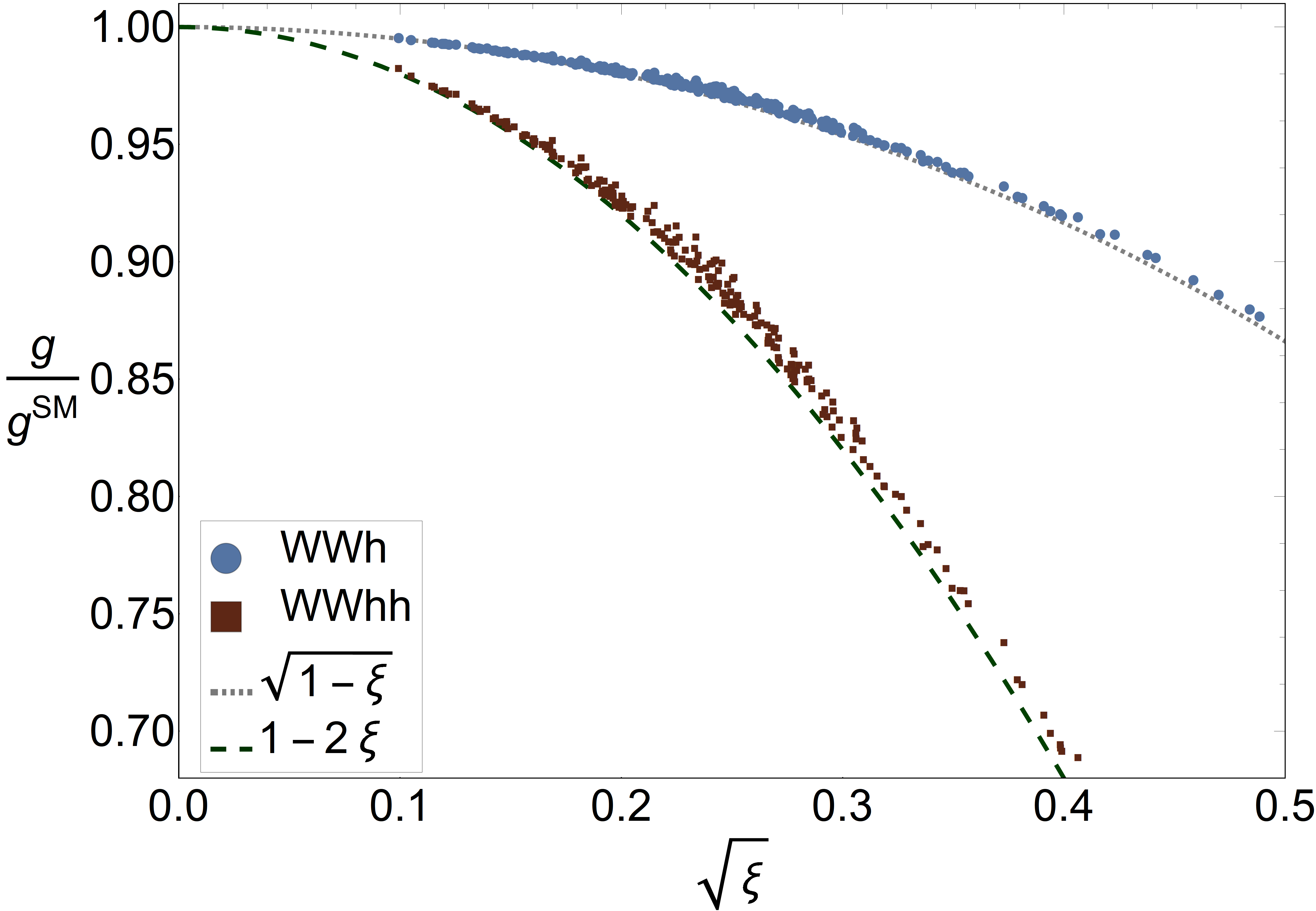}
\end{minipage}
\caption{On the left we show the Yukawa couplings of the top (in blue) and bottom (in red) quarks normalised to their SM values in the BP. The Yukawa coupling of the bottom quark was computed adding to the theory the Right-handed elementary bottom field, assuming that this addition does not change noticeably the potential. On the right we show the couplings of the $W$ boson with one and two Higgs bosons normalised to their SM values in the BP. The functions $\sqrt{1-\xi}$ and $1-2\xi$ are plotted in gray dotted line and green dashed line respectively.}
\label{F:yukTopBot}
\end{figure}

In Fig.~\ref{F:yukTopBot} we show the Yukawa couplings of the top and bottom quarks normalised to their SM values for all the BP. Both couplings show a clear suppression, which increases for bigger $\xi$. In the case of the bottom, the suppression can be well approximated by the leading order term of Eq.~(\ref{eq-yom2}): $\sqrt{1-\xi}$, since the corrections are proportional to $\sin^2\theta_d\ll 1$. On the other hand, there are points for which the top Yukawa shows a bigger suppression. These are due to non-negligible contributions from the second term of Eq.~(\ref{eq-yom2}), caused by the large mixing angles required by the high mass of the top quark.

The corrections to the Yukawa couplings of the light fermions are well approximated by the leading term, since the mixing of both chiralities are $\ll 1$. We also checked that approximation numerically.

The couplings between two $W$ bosons and one or two Higgs bosons were computed to leading order in Eq.~(\ref{eq-gom}). Corrections can be calculated, for example, by performing a perturbative diagonalisation of the mass matrix of $W$ resonances expanding in powers of $\xi$. We get:
\begin{align}
\frac{g_{WWh}}{g_{WWh}^{\rm SM}}\simeq& \sqrt{1-\xi}\left\{1+\xi\frac{3}{4}\frac{g_0^2}{(g_{0}^{2}+g_{1}^{2})^2}\frac{f^4}{f_{0}^{4}f_{1}^{2}}\left[f_{1}^{2}g_{1}^{2}+f_{0}^{2}\left(g_{0}^{2}+2g_{1}^{2}\right)\right]\right\}\,\nonumber\\
\frac{g_{WWhh}}{g_{WWhh}^{\rm SM}}\simeq& 1-2\xi+\xi(3-4\xi)\frac{g_0^2}{g_{0}^{2}+g_{1}^{2}}\frac{g_1^2f_{1}^{2}+f^2}{f_{0}^{2}+f_{1}^{2}} \ ,
\label{eq-gom2}
\end{align}
where $g_{WWh}^{\rm SM}$ and $g_{WWhh}^{\rm SM}$ are the couplings in the SM.

We show the numerical value of the ratios of Eq.~(\ref{eq-gom2}) for the BP in Fig.~\ref{F:yukTopBot}. The corrections are approximated with good accuracy by the terms depending just on $\xi$. This happens because the mixing between elementary and composite vector bosons is small. The couplings of the Higgs boson with the $Z$ boson are well approximated, to leading order, by the same function as the ones of the $W$.

The operator $hG_{\mu\nu}G^{\mu\nu}$ is generated at loop level by loops of SM quarks and their resonances. Each species of fermions contributes to the coefficient $c_g$ of this operator with:
\begin{equation}\label{eq-cg1}
c_g\propto \sum_n\frac{y_\psi^{(n)}}{m_\psi^{(n)}}A_{1/2}\left(\frac{m_h^2}{4m_\psi^{(n)2}}\right) \ ,
\end{equation}
with $A_{1/2}$ a loop function that can be found in App.~\ref{ap-tools}, normalised as $A_{1/2}(0)=4/3$. Taking into account that the resonances are much heavier than the Higgs, $c_g$ can be approximated by:
\begin{equation}\label{eq-cg2}
c_g\propto \frac{4}{3}\left[{\rm Tr}(Y_\psi M_\psi^{-1})-\frac{y_\psi^{(0)}}{m_\psi^{(0)}}\right]+\frac{y_\psi^{(0)}}{m_\psi^{(0)}}A_{1/2}\left(\frac{m_h^2}{4m_\psi^{(0)2}}\right) \ ,
\end{equation}
with $Y_\psi= \partial M_\psi/\partial\langle h\rangle$, and $M_\psi$ the mass matrix of the species $\psi$. Since $\lim_{\tau\to\infty}A_{1/2}(\tau)\to 0$, the last term of Eq.~(\ref{eq-cg2}) is suppressed for the light fermions. For the top, $A_{1/2}$ can be approximated by $4/3$, leading to $c_g\propto \frac{4}{3}{\rm Tr}(Y_t M_t^{-1})$. As has been extensively discussed in the literature~\cite{Azatov:2011qy,Falkowski:2007hz,Carena:2014ria}, ${\rm Tr}(Y_\psi M_\psi^{-1})=F_\psi/(\sqrt{\xi}f)$. For the light fermions and their composite partners, using Eq.~(\ref{eq-yom}) in (\ref{eq-cg2}), one obtains that their contribution to $c_g$ cancels to leading order. Thus the dominant contribution to the gluon coupling arise from the top sector, and the main corrections compared with the SM have the same factor as the Yukawa and gauge couplings: $\sqrt{1-\xi}$. 
 
The operator $hF_{\mu\nu}F^{\mu\nu}$ is also generated at loop level, this time by loops of charged fermions, vector bosons and the boson $\chi$. One can study the photon coupling $c_\gamma$ performing an analysis similar to $c_g$~\cite{Carena:2012xa}. Since the Yukawa and the $W$ couplings are corrected by the same factor, $c_\gamma$ is suppressed to leading order by a factor $\sqrt{1-\xi}$ compared with the SM. There are also smaller corrections arising from the fact that there are new states running in the loop. The contribution of the $\chi$ mediated diagrams to the total amplitude was found to be around 3 to 4 orders of magnitude smaller than the other contributions and therefore was not included in the subsequent computations.

Other interesting interaction induced at one-loop level is $hZ\gamma$. The presence of the symmetry $P_{LR}$ in the theory avoids large effects in this process~\cite{Azatov:2011qy}. Following an analysis similar to $c_g$ and $c_\gamma$~\cite{Carena:2014ria}, one can obtain that the main correction to $c_{hZ\gamma}$, compared with the SM, is given by a suppression factor $\sqrt{1-\xi}$.

\begin{figure}[h!]
\begin{center}
\includegraphics[width=\textwidth]{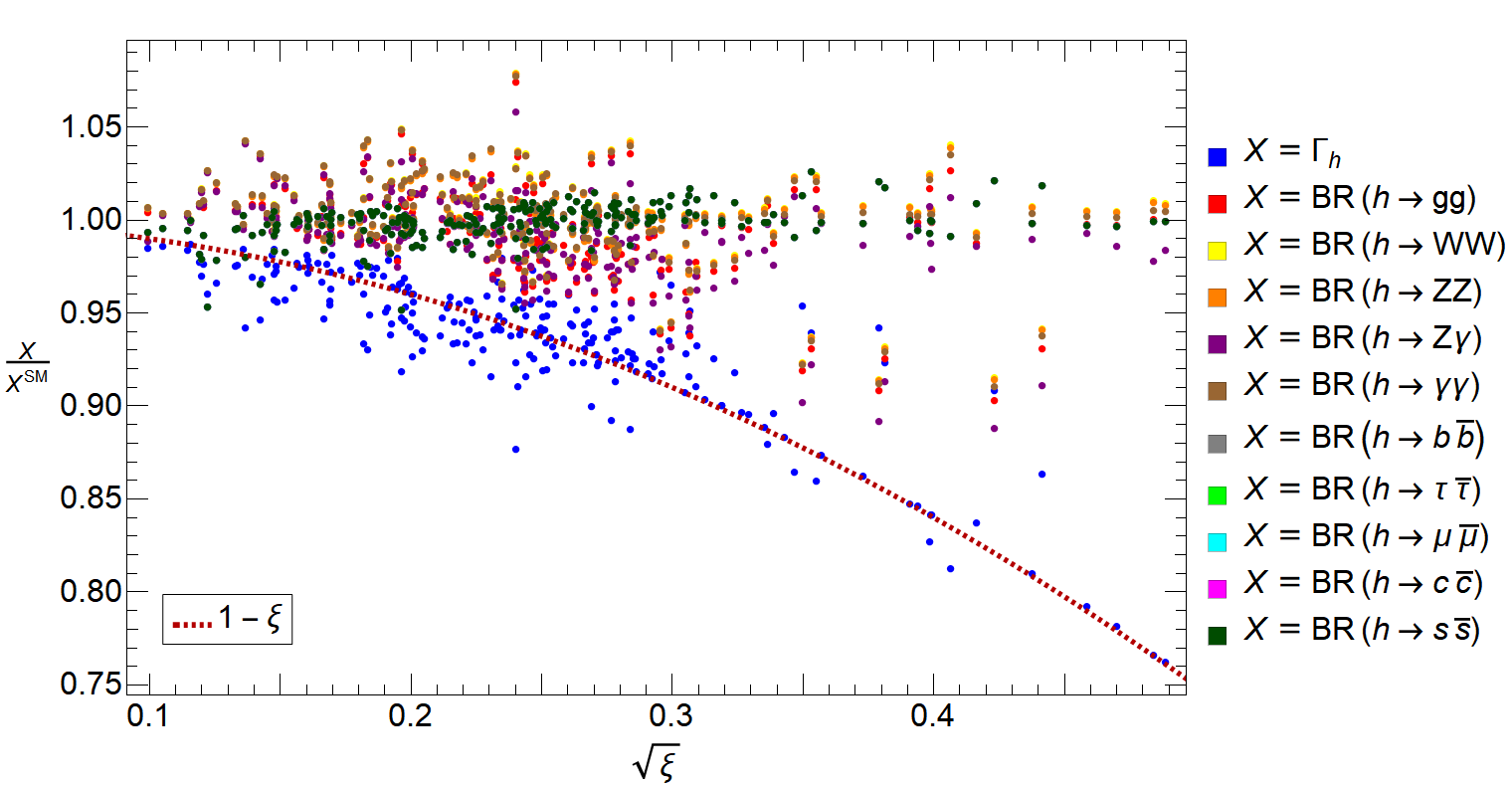}
\caption{Total width and branching ratios of the Higgs boson normalised to their SM values in the BP. The dotted line is $1-\xi$.}
\label{F:WidthBRHiggs}
\end{center}
\end{figure}

Armed with the previous analysis of Higgs couplings, we have computed the total width, branching ratios and production cross sections of the Higgs boson, considering both the corrections to couplings with the SM particles and the influence of new resonances in loop-level processes. The branching ratios and the total width normalised to the SM for the BP are shown in Fig.~\ref{F:WidthBRHiggs}. The total width is suppressed with respect to the SM by an approximate factor $(1-\xi)$, since all the couplings are approximately suppressed by $\sqrt{1-\xi}$. The branching ratios show smaller corrections without a clear trend towards rising or decreasing, as expected, since the main correction cancels in the ratio. For $\sqrt{\xi}<0.2$, where the constraints for the resonance masses are satisfied, the total width is suppressed in no more than $10\%$ and the branching ratios are corrected in less than $5\%$.

\begin{figure}[h!]
\begin{center}
\begin{minipage}[b]{0.49\textwidth}
\includegraphics[width=\textwidth]{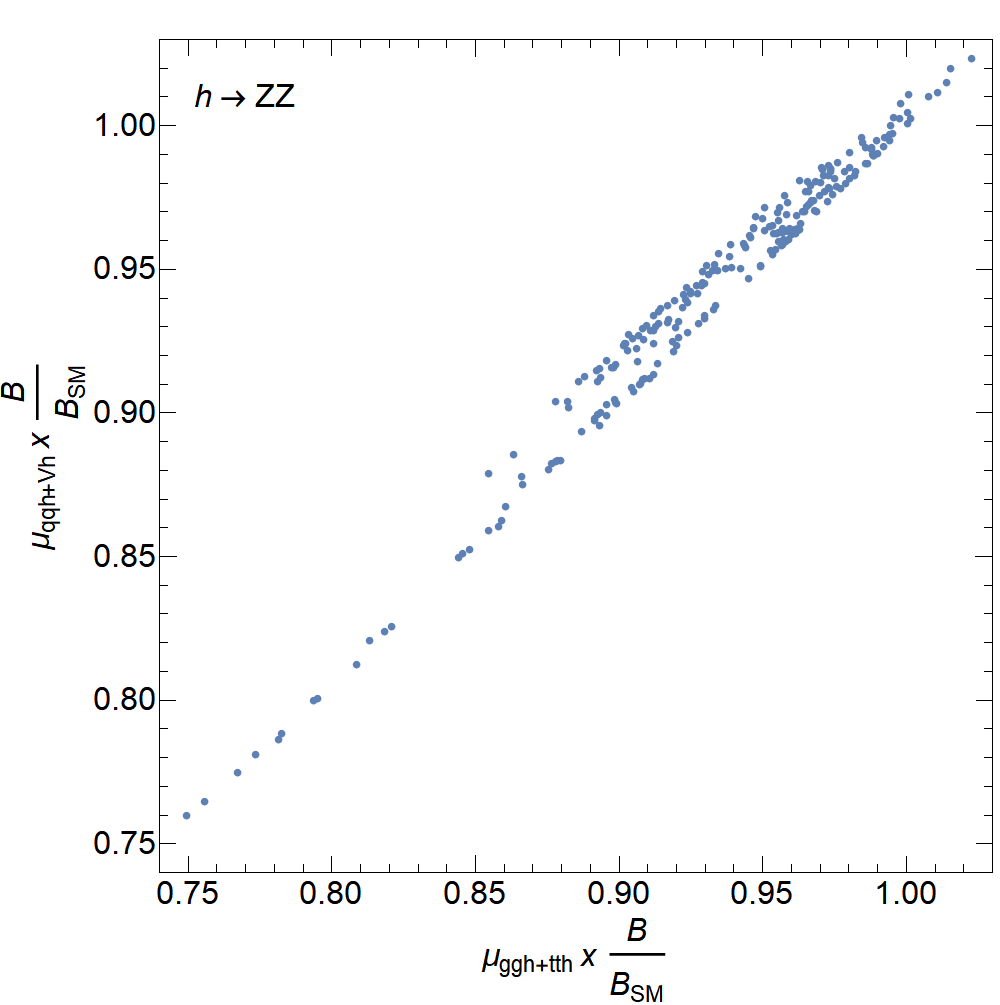}
\end{minipage}\hfill
\begin{minipage}[b]{0.49\textwidth}
\includegraphics[width=\textwidth]{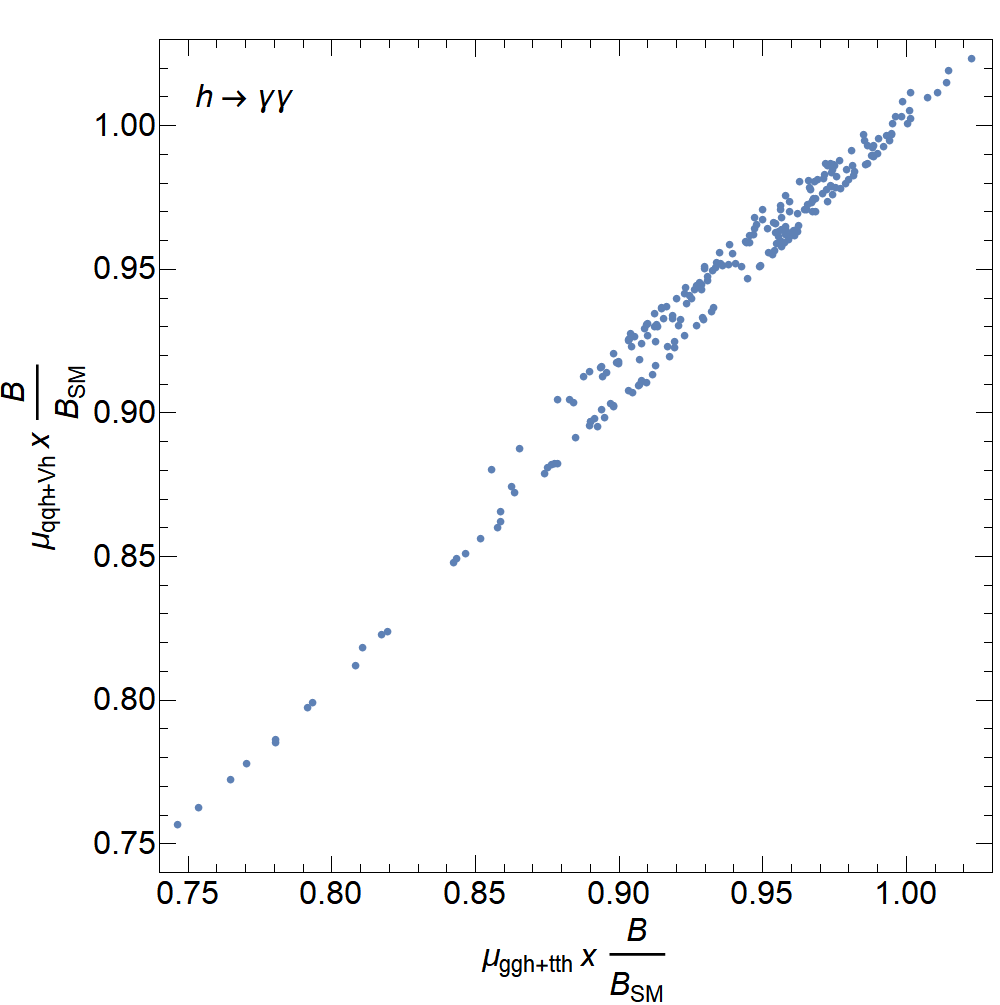}
\end{minipage}
\caption{Left panel: Rates in the $h\rightarrow ZZ$ decay channel separated according to production mode: gluon fusion ($+t\overline{t}h$) versus VBF ($+hW/Z$). Right panel: Same for the $h\rightarrow\gamma\gamma$. The points correspond to the BP.}
\label{F:MUHIGGS}
\end{center}
\end{figure}

We have also considered the total cross sections for the $ZZ$ and $\gamma\gamma$ decay modes, separating the production modes: gluon fusion ($+t\overline{t}h$) and VBF ($+hW/Z$), as done by the ATLAS and CMS collaborations and in previous works\cite{ATLAS:2012klq,ATLAS:2013oma,Chatrchyan:2013mxa,Carena:2014ria}. Defining the production signal strength as $\mu_i=\sigma^{\rm Model}(i)/\sigma^{\rm SM}(i)$, we show our results in Fig.~\ref{F:MUHIGGS}. Since the branching ratios are just slightly modified with respect to the SM and all the couplings and diagrams involved in the production modes are suppressed similarly, both plots show points not far from a straight line tilted $45^{o}$. The points which are slightly over such straight line correspond to the region where the top Yukawa coupling is additionally suppressed due to very high mixing. As expected, the results are very similar to the MCHM$_{5-10-10}$.

Finally, we studied the Higgs boson self-interactions. Defining the Higgs cubic and quartic self-interactions as $V\supset \lambda_{3h} h^3+ \lambda_{4h} h^4$, in the SM these couplings are given by: $\lambda_{3h}^{SM}=\frac{m_{h}^{2}}{2 v}$ and $\lambda_{4h}^{SM}=\frac{m_{h}^{2}}{8 v^2}$, with $m_h$ the physical Higgs mass. By making use of Eq.~(\ref{eq-vsin4}), one can compute $\lambda_{3h}$ and $\lambda_{4h}$ in the present model. After minimization of the potential, we obtain:
\begin{equation}
\lambda_{3h}=\lambda_{3h}^{SM}\left(1-\frac{3}{2}\xi+\mathcal{O}\left(\xi^2\right)\right), 
\qquad
\lambda_{4h}=\lambda_{4h}^{SM}\left(1-\frac{25}{3}\xi+\mathcal{O}\left(\xi^2\right)\right),
\label{E:HiggsSelfCoupApprox}  
\end{equation}
These results tell us that the Higgs self couplings are expected to be suppressed with respect to their Standard Model values. 

\begin{figure}[h!]
\begin{center}
\includegraphics[width=0.9\textwidth]{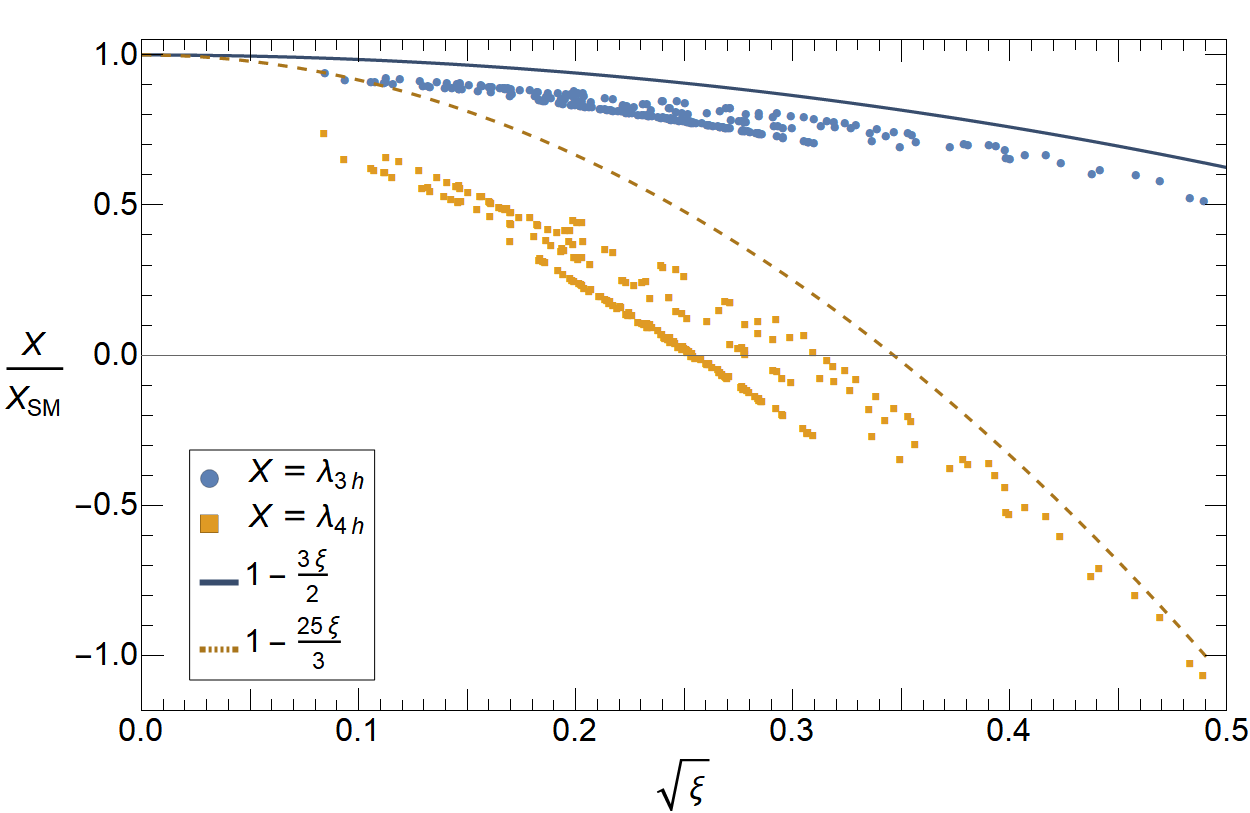}
\caption{Trilinear ($\lambda_{3h}$, blue circles) and quartic ($\lambda_{4h}$, orange squares) Higgs self-couplings for the BP of our model. We show the analytical approximations for them: in blue continuous line for the trilinear self-coupling, and in orange dashed line for the quartic self-coupling.}
\label{F:selfIntHiggs}
\end{center}
\end{figure}

We also computed the numerical value of the Higgs self couplings in the BP of our model. The result is plotted in Fig.~\ref{F:selfIntHiggs}, where we have included the approximations of Eq.~\ref{E:HiggsSelfCoupApprox} for comparison. Both self couplings are more suppressed with respect to the Standard Model values than the analytical estimates. When the quartic coupling is negative, terms in the potential with higher powers of $h$ must be positive to make it grow around the minimum. Given that the suppression of both self couplings is bigger than 10\% with respect to the Standard Model values even for $\sqrt{\xi}\sim 0.1$, their study is a interesting probe for this class of models.

\subsection{A stable exotic state}
The $\chi$ boson is heavier than all the SM particles and, being a pNGB, it is in general the lightest resonance (with the exception of the Higgs). Being $\chi$ a singlet of SU(3)$_c$ with electric charge $Q=2/3$, it can not decay to SM states. Therefore it is in general stable. 

The stability of $\chi$ has been used in Ref.~\cite{Balkin:2017aep}, where U(1)$_X$ was not included in SO(7), to make it a dark matter candidate. In that case its stability was associated to its charge under a conserved global U(1). In the present case that U(1) has been gauged, according to Eq.~(\ref{eq-Y}).

Despite $\chi$ being a pNGB, there are regions of the parameter space of our model where there are fermion resonances $\Psi^{(n)}$ with exotic charges and lighter than $\chi$ (as color triplets with vanishing or integer electric charges), see Fig.~\ref{F:masResVecPI}. In this case, the pNGB $\chi$ could decay to $\Psi^{(n)}$ plus SM particles. The lightest fermion with exotic charges can not decay to SM particles.

Thus, in the present model one can expect the presence of a stable resonance with exotic charges. In large regions of the parameter space this state is the colorless scalar $\chi$, but in other regions we find it to be a fermion electromagnetically neutral transforming in the 
fundamental representation of SU(3)$_c$. This neutral quark, as well as neutral octets, has been called {\it quorn} in the literature, with hadrons $\Psi^{(n)}\bar\Psi^{(n)}$ being considered as good candidates for dark matter~\cite{DeLuca:2018mzn,Gross:2018zha}. Ref.~\cite{DeLuca:2018mzn} has shown that the abundance of hybrid hadrons, made of SM quarks and $\Psi^{(n)}$, is suppressed by several orders of magnitude compared with the abundance of a $\Psi^{(n)}\Psi^{(n)}\Psi^{(n)}$ hadron state. Hadrons containing $\Psi^{(n)}$ and charged fermionic resonances $\Psi^{(m)}_{Q}$, as $\Psi^{(n)}\Psi^{(n)}\Psi^{(m)}_{Q}$ or $\Psi^{(n)}\Psi^{(m)}_Q\Psi^{(m')}_{Q'}$, could also lead to charged hadronic states. However $\Psi^{(m)}_{Q}$ can decay to SM quarks or to $\Psi^{(n)}\bar qq'$ through four fermion operators, leading to either hybrid hadrons or pure $\Psi^{(n)}$ hadrons. A calculation of the abundance of electrically charged stable hadrons (that must satisfy very stringent bounds), as well as the abundance of neutral hadrons made of $\Psi^{(n)}$ only, as function of the parameters of the theory, must be done to know the viability of the present model. This is a rather complicated calculation, that is  beyond the scope of the present work, and is left for the future.

By changing the representation of the quarks under SO(7) it is possible to obtain colored states with fractional charges only. An example is given by the representation {\bf 8} of SO(7), that under SO(4)$\times$U(1)$_X$ decomposes as: ${\bf 8}\sim({\bf2},{\bf1})_{\pm 1/(2\sqrt{2})}+({\bf1},{\bf2})_{\pm 1/(2\sqrt{2})}$. By taking $\alpha = \sqrt{2}/3$ in Eq.~(\ref{eq-Y}), the quark doublet $q$ can be identified with $({\bf2},{\bf1})_{1/(2\sqrt{2})}$, whereas the singlets $u$ and $d$ can be identifed with the components of $({\bf1},{\bf2})_{1/(2\sqrt{2})}$ with $T^3_R=1/2$ and $-1/2$, respectively. In this case, the other components of the {\bf 8} have electric charges $Q=1/3$ and -2/3, whereas the boson $\chi$ has $Q=1/3$.

\subsection{Phenomenology of the new charged scalar}
The new charged scalar boson, $\chi$, has a higher mass than the Higgs without an additional tuning of the parameters, as shown in Fig.~\ref{F:spectrum}, ~\ref{F:cnIntAngUQdetail} and ~\ref{F:cvIntf0g1}. More precisely, when the Higgs mass is near its measured value, $m_\chi\gtrsim 500$ GeV. As for the Higgs, we studied the correlation between $m_\chi$ and $\xi$, finding that for $\sqrt{\xi}\lesssim 0.2$, $m_\chi$ is in general $\gtrsim 1$~TeV.

It is easy to notice the similarity between the $\chi$ boson here presented and an spartner of an up-type quark included in SUSY models, such as the stop. The main difference is that the $\chi$ boson does not interact strongly, which has important consequences over its stability and production.

\subsubsection{Production and signatures at LHC}

At an hadron collider a pair of $\chi$ bosons can be produced from gluons only at loop level, while a pair of squarks can be produced from gluons at tree level. Furthermore, the symmetries of the theory require at least one resonance in the initial state in order to produce a single $\chi$ boson.

\begin{figure}[h!]
\begin{center}
\begin{minipage}{0.4\textwidth}
\includegraphics[width=\textwidth]{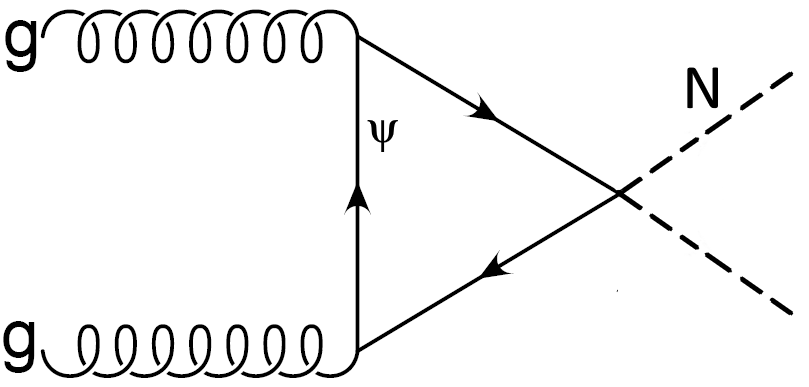}
\vskip1cm
\end{minipage}
\begin{minipage}{0.58\textwidth}
\includegraphics[width=\textwidth]{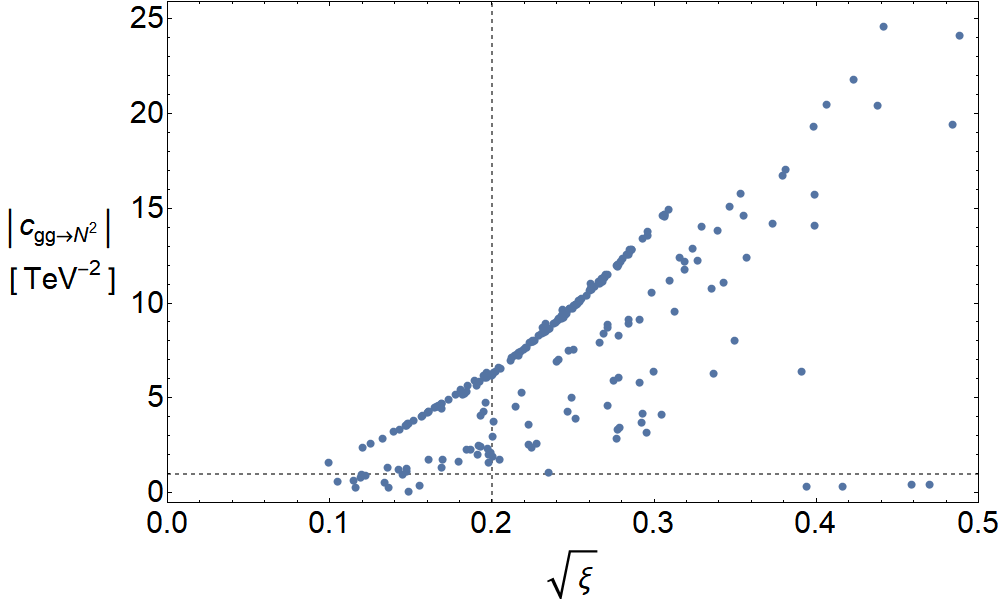}
\end{minipage}
\caption{Left panel: Feynman diagram for the main production channel of $\chi$ boson pairs in an hadron collider. $\Psi$ is any fermion, elementary or composite, with mass $m_\psi^{(n)}$. The amplitude of the depicted diagram is proportional to the coefficient $c_{gg\chi^2}$ defined in Eq.~\ref{E:gluonFusN2coef}. Right panel: absolute value of $c_{gg\chi^2}$ as function of $\sqrt{\xi}$, for the BP. The dotted horizontal line is at 1 TeV$^{-2}$ and the dotted vertical line is at $\sqrt{\xi}=0.2$.}
\label{F:FeynDiagProd2N}
\end{center}
\end{figure}

An example of the kind of processes that allow pair production of $\chi$ at hadron collider is shown in Fig.~\ref{F:FeynDiagProd2N}. Since $\chi$ is charged under U(1)$_Y$, it is also possible to produce it in pairs by $q\bar q\to\gamma,Z\to\chi\chi^*$, as well as by production of a single Higgs off-shell that decays to $\chi\chi^*$, see Eq.~(\ref{E:potO4}). Below we analyse the gluon fusion interaction of Fig.~\ref{F:FeynDiagProd2N}. 

The fermion loop includes all the fermions, whether elementary or composite. The amplitude of the Feynman diagram in Fig.~\ref{F:FeynDiagProd2N} is proportional to the following coefficient:
\begin{equation}
c_{gg\chi^2}=\sum_{\psi,n}\frac{y^{(n)}_{\psi\chi^2}}{m_\psi^{(n)}}A_{1/2}\left(\frac{Q^2}{(2m^{(n)}_\psi)^2}\right) \ ,
\label{E:gluonFusN2coef}
\end{equation}
where the sum runs over all the fermions included in the theory. The function $A_{1/2}$ is the same loop function usually found in the gluon fusion computations (see App.~\ref{ap-tools}), $Q^2$ is the square of the total 4-momentum of the final state $\chi\chi^*$ ($Q^2\geq 4m_{\chi}^2$ for on-shell production) and $y^{(n)}_{\psi\chi^2}$ is the coupling between the n-th fermion and a pair of $\chi$ bosons: $\mathcal{L}\supset y^{(n)}_{\psi\chi^2} |\chi|^2 \overline{\Psi}_{n}\Psi_{n}$. These couplings can be calculated in the 2-site theory by expanding the Yukawa interactions at site-1 to second order in $\chi$, and then rotating to the mass basis.

We have computed $c_{gg\chi^2}$ for the BP of our scan, fixing $Q^2=4m_\chi^2$. By analysing the contributions of the fermions with different charges, we find that, in general, the sector of neutral resonances dominate by a factor of order two over those with charge $-1$ and $+2/3$, that are in the second and third place, respectively. The different role of the fermions, when compared with $c_g$ in Higgs production, is due to several reasons. First, all the SM fermions, even the top quark, are much lighter than $2m_\chi$, thus $A_{1/2}$ is relatively small and equal for all of them. On the other hand, $|A_{1/2}(\tau)|$ has a maximum for $\tau\sim 1$, thus $c_{gg\chi^2}$ receives larger contributions from fermion resonances with masses of order $m_\chi$\footnote{The function $|A_{1/2}(\tau)|$ has a maximum at $\tau\cong 1.473$ and its value there is $\cong 2.419$, while its limit when $\tau\rightarrow 0^+$ is $\frac{4}{3}$.}. Finally, since the Yukawa couplings do not show large differences between different species of fermions, and the neutral resonances have the largest multiplicity, see Table~\ref{T:especResFerm}, one can expect a larger contribution from them.

We show the predictions for $c_{gg\chi^2}$, for the BP of the scan, with $Q^2=4m_\chi^2$, on the right panel of Fig~\ref{F:FeynDiagProd2N}. For $\sqrt{\xi}<0.2$ the coupling is of order 1-6~TeV$^{-2}$, increasing with $\xi$.

In order to give a complete analysis of the possible signatures of the $\chi$ boson at the LHC, a careful computation of the different contributions to pair-production, as well as the possible final signatures, is needed. These tasks are beyond the scope of this work.

\section{Discussions and conclusions}\label{sec-conclusions}
We have studied the unification of the EW and custodial symmetry of the SM into a simple global group of a new composite sector, with the Higgs arising as a composite pNGB. The smallest coset with these properties is SO(7)/SO(6), generating the Higgs as well as a new pNGB $\chi$, that is a color- and SU(2)$_L$-singlet and has hypercharge $2/3$. We have found that fermion resonances in the representations ${\bf 21}$ and ${\bf 35}$ of SO(7) can mix with the SM quark doublets and singlets, respectively, whereas those in ${\bf 7}$ and ${\bf 21}$ can mix with the SM lepton doublets and singlets. 

We have presented a 2-site theory that allows to describe the lowest lying level of resonances of the composite sector. We have studied the spectrum of resonances, and we have obtained the low-energy effective theory resulting from the integration of the massive resonances. We have computed the potential generated at one-loop by the interactions of the composite sector with the elementary fermions and gauge SM fields, that explicitly break the SO(7) symmetry of the composite sector. We have studied some approximations of the potential, and we have found regions of the parameter space that can reproduce the masses of the EW gauge bosons, as well as the top quark and the Higgs, with $f\sim 1.2$~TeV. The mass of the new pNGB is of order 1~TeV.

The embedding of the EW symmetry into SO(7) has led to a set of new resonances, compared with the MCHM, like colored fermions with integer electric charges, as well as spin-one states with charges $|Q|=1/3,2/3,5/3$. The colorless pNGB with $Q=2/3$ and a set of fermions are the lightest states, with one of them being stable. Thus the model gives a very rich spectrum and new phenomenology at colliders, compared with the usual scenarios of composite Higgs. The understanding of this phenomenology requires several analysis that have not been done yet. As an example, although we have studied the size of the coupling that allows pair creation of $\chi$ in gluon fusion, a careful studied of its production cross-section at LHC is required. That analysis is beyond the scope of this work and has been left for the future. The exotic fermion resonances could also be produced in pairs, at hadronic colliders as LHC, by QCD interactions. They are expected to hadronize into mesons $\Psi^{(n)}\bar q$ or baryons $\Psi^{(n)}qq'$, that could be charged and leave traces in the detectors. Searches from experiments at LHC give bounds of order $m_\psi^{(n)}\gtrsim 2$~TeV~\cite{ATLAS:2018yey,CMS:2016ybj,DeLuca:2018mzn}, however a dedicated study of their production and detection, as well as a recasting of existing analysis, must also be done.

The presence of an exotic stable state can give a serious problem with cosmology, that could made the present model not viable. A necessary condition is that the abundance of hybrid hadrons made of $\Psi^{(n)}$ and SM quarks does not modify the cosmology. On the other hand, baryons made of $\Psi^{(n)}$ only could eventually give a fraction of dark matter~\cite{DeLuca:2018mzn}. A careful calculation of the cosmological evolution of these quantities must be done.

Another interesting topic on composite Higgs models with partial compositeness are the bounds from CP-violating dipole operators, since they give some of the strongest constraints. In flavor anarchic theories, the electromagnetic dipole moment of the quarks give strong bounds, pushing $f$ up to $\gtrsim 5$~TeV~\cite{Konig:2014iqa}. Ref.~\cite{Antipin:2014mda} has shown that in the MCHM one can generically expect also contributions from the non-linearities of the theory, associated to the non-linear realization of the symmetries, as well as UV contributions from the composite sector. Unification of the composite EW-symmetry allow the possibility of a cancellation between different contributions to the electromagnetic dipole operators.~\footnote{See also Ref.~\cite{Panico:2018hal} for an analysis of the constraints from the results of the ACME collaboration~\cite{Andreev:2018ayy}.} We have computed, in the effective theory, the invariants compatible with the coset SO(7)/SO(6) that lead to dipole operators:
\begin{equation}
\Gamma_{\bf rst}(p^2)\left[\overline{(U^\dagger\psi_L)}_{\bf r}(U^\dagger a_{\mu\nu})_{\bf s}\sigma^{\mu\nu}(U^\dagger\psi_R)_{\bf t}\right]_{\bf 1}
\end{equation}
where $\Gamma_{\bf rst}(p^2)$ are momentum dependent form-factors that codify the composite dynamics, they are independent of the NGBs that are contained in the matrices $U$, $a_{\mu\nu}$ is the SO(7) field strength and the subindices ${\bf r},{\bf s},{\bf t}$ label SO(6) irreducible representations. The product ${\bf r}\otimes {\bf s}\otimes{\bf t}$ is projected onto the SO(6) singlet ${\bf 1}$. Putting the Higgs to its vev, and keeping only the dynamical elementary fermions and gauge fields, we find that, for $q_L$ embedded in ${\bf 21}$, $u_R$ embedded in ${\bf 7}$ or ${\bf 35}$ and $d_R$ embedded in ${\bf 35}$, the electromagnetic dipole operator of up- and down-type fermions does not vanish for generic form-factors $\Gamma$. Thus for these representations the bounds from EDMs are not relaxed.

As mentioned in the previous paragraph, although we have not studied it in detail, composite partners of $u_R$ could also be embedded in a ${\bf 7}$. In this case one can expect some modifications in the analysis of the potential, as well as in the spectrum of fermion resonances. The study of this, as well as other representations for which the electromagnetic dipole could be suppressed, is an interesting avenue for future work.

\section*{Acknowledgements}
We thank Christophe Grojean for his comments on an early version of the manuscript and on the $\chi$ mediated Higgs decay to photons and Giuliano Panico for discussions on the Higgs trilinear coupling. We also thank Jesse Thaler for a discussion about coloured relic densities. L. D.'s work is partially supported by Argentinian ANPCyT PICT 2013-2266. L. D. thanks ICAS-UNSAM for hospitality during part of this work. A. R.'s work was partially funded by CNEA-Instituto Balseiro Master's scholarship. A. R. thanks  DESY Theory Group for support during the final stages of this work.

\appendix
\section{The group SO(7)}\label{ap-so7}
A basis for the algebra of SO(7) in the fundamental representation can be written as:
\begin{equation}
(T_{ij})_{k\ell}=\frac{i}{\sqrt{2}} (\delta_{ik}\delta_{j\ell}-\delta_{i\ell}\delta_{jk}) \ , 
\qquad i<j, \ ,i=1,\dots 6 \ ,j=2,\dots 7 \ ,
\end{equation}
with the normalization ${\rm tr}(T_{ij}T_{mn})=\delta_{im}\delta_{jn}$.
The generators of SU(2)$_\times$SU(2)$_R\times$U(1)$_X$ are given by:
\begin{align}
T_{1}^{L}&=-\frac{1}{\sqrt{2}}\left(T_{23}+T_{14}\right)\ , \qquad
T_{2}^{L}&=\frac{1}{\sqrt{2}}\left(T_{13}-T_{24}\right)\ , \qquad
T_{3}^{L}&=-\frac{1}{\sqrt{2}}\left(T_{12}+T_{34}\right)\ , \nonumber\\
T_{1}^{R}&=-\frac{1}{\sqrt{2}}\left(T_{23}-T_{14}\right)\ , \qquad
T_{2}^{R}&=\frac{1}{\sqrt{2}}\left(T_{13}+T_{24}\right)\ , \qquad
T_{3}^{R}&=-\frac{1}{\sqrt{2}}\left(T_{12}-T_{34}\right)\ , \nonumber\\
X&=T_{67}\ . & &
\label{E:GenSU2SU2U1}
\end{align}

The adjoint representation, ${\bf 21}$, can be obtained from the structure constants. Representation ${\bf 35}$ is obtained from $\mathbf{7}\otimes\mathbf{21}\sim\mathbf{7}\oplus\mathbf{35}\oplus\mathbf{105}$.

\section{Mass matrices and form-factors in the 2-site theory}\label{ap-2site}
Below we show the mass matrices in the vacuum of Eq.~(\ref{eq-vev}). For the neutral fermions:
\begin{equation}
M_{f,0}=\left(\begin{array}{c|c|c}
m_Q I_{5\times5}  & Y_{0}^{u}&Y_{0}^{d}\\
\mathbf{0}_{7\times5}&m_U I_{7\times7}&\mathbf{0}_{7\times7}\\
\mathbf{0}_{7\times5}&\mathbf{0}_{7\times7}&m_D I_{7\times7}\\
\end{array}\right),
\end{equation}
with:
\begin{equation}
Y_{0}=\left(\begin{array}{ccccccc}
 0 & 0 & 0 & 0 & -\frac{1}{2} \alpha_1 & -\frac{1}{2} \alpha_1 & 0 \\
 0 & 0 & 0 & 0 & -\frac{1}{2} \alpha_1 & -\frac{1}{2} \alpha_1 & 0 \\
 -\frac{1}{2} \alpha_1 & \frac{1}{2} \alpha_1 & \frac{1}{2} \alpha_1 & -\frac{1}{2} \alpha_1 & 0 & 0 & -\alpha_2 \\
\frac{\alpha_1}{2 \sqrt{2}} & \frac{\alpha_1}{2 \sqrt{2}} & \frac{\alpha_1}{2 \sqrt{2}} &\frac{\alpha_1}{2 \sqrt{2}} & -\alpha_2 & 0 & 0 \\
-\frac{\alpha_1}{2 \sqrt{2}} & -\frac{\alpha_1}{2 \sqrt{2}} & -\frac{\alpha_1}{2 \sqrt{2}} &-\frac{\alpha_1}{2 \sqrt{2}} & 0 & -\alpha_2 & 0 \end{array}\right),
\end{equation}
and\begin{equation}
\alpha_1= i f_{1} y_{\Psi} \sqrt{\xi}
\ , \qquad
\alpha_2= f_{1} y_{\Psi} \sqrt{1-\xi}\ .
\label{E:defApend5}
\end{equation}
For down-type fermions:
\begin{equation}
M_{f,-\frac{1}{3}}=\left(
\begin{array}{c|c|c|c}
 m_Q & \gamma_1\left(y_{U}\right)& \gamma_1\left(y_{D}\right)&0\\
\mathbf{0}_{3\times1} & m_U I_{3\times3} & \mathbf{0}_{3\times3} & \mathbf{0}_{3\times1} \\
\mathbf{0}_{3\times1} & \mathbf{0}_{3\times3} & m_D I_{3\times3} & \mathbf{b} \\
 -f_{0} \lambda_{q} & \mathbf{0}_{1\times3} & \mathbf{0}_{1\times3} &0 \end{array}
\right),
\end{equation}
with
\begin{equation} 
\gamma_1\left(x\right)=\left(\begin{array}{ccc}
\frac{f_{1}  x \sqrt{\xi}}{\sqrt{2}} & -\frac{f_{1} x \sqrt{\xi}}{\sqrt{2}} & -f_{1}  x \sqrt{1-\xi}
\end{array}\right)\ ,
\qquad
\mathbf{b}=\left(\begin{array}{c}
0\\
-f_{0} \lambda_{d}^{*}\\
0
\end{array}\right)\ .
\end{equation}
For $Q=1/3$
\begin{equation}
M_{f,\frac{1}{3}}=\left(
\begin{array}{c|c|c}
 m_Q & \gamma_2\left(y_{U}\right)& \gamma_2\left(y_{D}\right)\\
\mathbf{0}_{3\times1} & m_U I_{3\times3} & \mathbf{0}_{3\times3}\\
\mathbf{0}_{3\times1} & \mathbf{0}_{3\times3} & m_D I_{3\times3}\end{array}
\right),
\end{equation}
with:
\begin{equation} 
\gamma_2\left(x\right)=\left(\begin{array}{ccc}
-\frac{f_{1}  x \sqrt{\xi}}{\sqrt{2}} & \frac{f_{1} x \sqrt{\xi}}{\sqrt{2}} & -f_{1}  x \sqrt{1-\xi}
\end{array}\right)\ ,
\end{equation}
For $Q=-2/3$:
\begin{equation}
M_{f,-\frac{2}{3}}=\left(
\begin{array}{c|c|c}
 m_Q I_{3\times3} &\beta_1\left(y_{U}\right)&\beta_1\left(y_{D}\right)\\
\mathbf{0}_{4\times3} & m_U I_{4\times4} & \mathbf{0}_{4\times4}\\
\mathbf{0}_{4\times3} & \mathbf{0}_{4\times4} & m_D I_{4\times4}\end{array}
\right),
\end{equation}
with:
\begin{equation}
\beta_1\left(x\right)=\left(
\begin{array}{cccc}
 0 & 0 & \frac{i f_{1} x \sqrt{\xi}}{\sqrt{2}} & -\frac{i f_{1} x \sqrt{\xi}}{\sqrt{2}}
   \\
 -\frac{1}{2} f_{1} x \sqrt{\xi} & \frac{1}{2} f_{1} x \sqrt{\xi} & -f_{1} x \sqrt{1-\xi} & 0 \\
 -\frac{1}{2} f_{1} x \sqrt{\xi} & \frac{1}{2} f_{1} x \sqrt{\xi} & 0 & -f_{1} x
   \sqrt{1-\xi} \\
\end{array}
\right).
\end{equation}
For up-type quarks:
\begin{equation}
M_{f,\frac{2}{3}}=\left(
\begin{array}{c|c|c|c}
 m_Q I_{3\times3} & \beta_2\left(y_{U}\right)& \beta_2\left(y_{D}\right)&\mathbf{0}_{3\times1}\\
\mathbf{0}_{4\times3} & m_U I_{4\times4} & \mathbf{0}_{4\times4} & \mathbf{g} \\
\mathbf{0}_{4\times3} & \mathbf{0}_{4\times4} & m_D I_{4\times4} & \mathbf{0}_{4\times1} \\
\mathbf{w} & \mathbf{0}_{1\times4} & \mathbf{0}_{1\times4} &0 \end{array}
\right),
\end{equation}
with:
\begin{align}
&\beta_2\left(x\right)=\left(
\begin{array}{cccc}
 0 & 0 & -\frac{i f_{1} x \sqrt{\xi}}{\sqrt{2}} & \frac{i f_{1} x \sqrt{\xi}}{\sqrt{2}}
   \\
 \frac{1}{2} f_{1} x \sqrt{\xi} & -\frac{1}{2} f_{1} x \sqrt{\xi} & -f_{1} x \sqrt{1-\xi} & 0 \\
 \frac{1}{2} f_{1} x \sqrt{\xi} & -\frac{1}{2} f_{1} x \sqrt{\xi} & 0 & -f_{1} x
   \sqrt{1-\xi}
\end{array}
\right)\ ,
\qquad
\mathbf{g}=\left(\begin{array}{c}
0\\
-f_{0} \lambda_{u}^{*}\\
0\\
0
\end{array}\right)\ ,
\nonumber \\
&\mathbf{w}=\left(\begin{array}{ccc}
0\,&0\,&
-f_{0} \lambda_{q}
\end{array}\right)\ .
\end{align}
For $Q=1$:
\begin{equation}
M_{f,1}=\left(
\begin{array}{c|c|c}
 m_Q I_{3\times3} & \epsilon\left(y_{U}\right)& \epsilon\left(y_{D}\right)\\
\mathbf{0}_{4\times3} & m_U I_{4\times4} & \mathbf{0}_{4\times4}\\
\mathbf{0}_{4\times3} & \mathbf{0}_{4\times4} & m_D I_{4\times4}\end{array}
\right),
\end{equation}
where:
\begin{equation}
\epsilon\left(x\right)=\left(
\begin{array}{cccc}
 0 & \frac{i f_{1} x \sqrt{\xi}}{\sqrt{2}} & 0 & \frac{i f_{1} x \sqrt{\xi}}{\sqrt{2}}\\
 \frac{1}{2} i f_{1} x \sqrt{\xi} & -f_{1} x \sqrt{1-\xi} & \frac{1}{2} i f_{1} x 
\sqrt{\xi} & 0 \\
 -\frac{1}{2} i f_{1} x \sqrt{\xi} & 0 & -\frac{1}{2} i f_{1} x \sqrt{\xi} & -f_{1}
   x \sqrt{1-\xi} \\
\end{array}
\right) \ .
\end{equation}
$M_{f,-1}$ is obtained from $M_{f,1}$ by changing the sign of the coefficients of the non-diagonal elements of the first line.
For $Q=\mp 5/3$:
\begin{equation}
M_{f,\pm\frac{5}{3}}=\left(
\begin{array}{c|c|c}
 m_Q & \gamma_{1,2}\left(y_{U}\right)& \gamma_{1,2}\left(y_{D}\right)\\
\mathbf{0}_{3\times1} & m_U I_{3\times3} & \mathbf{0}_{3\times3}\\
\mathbf{0}_{3\times1} & \mathbf{0}_{3\times3} & m_D I_{3\times3}\end{array}
\right) \ .
\end{equation}

The case without $d_R$ resonances can be obtained by taking $\lambda_d=0$ and keeping only the blocks of the matrices involving $q_L$ and $u_R$, as well as $Q$ and $U$ resonances.

The fermion form-factors are:
\begin{align}
\Pi_{6}^{q}\left(p^{2}\right) & =\frac{f_{0}^{2}\left|\lambda_{q}\right|^{2}}{m_{Q}^{2}-p^{2}},\\
\Pi_{15}^{q}\left(p^{2}\right) & =\frac{f_{0}^{2}\left|\lambda_{q}\right|^{2}\left(f_{1}^{2}\left|y_{D}\right|^{2}\left(m_{U}^{2}-p^{2}\right)+f_{1}^{2}\left|y_{U}\right|^{2}\left(m_{D}^{2}-p^{2}\right)+\left(m_{D}^{2}-p^{2}\right)\left(m_{U}^{2}-p^{2}\right)\right)}{\left(m_{Q}^{2}-p^{2}\right)\left(m_{U}^{2}-p^{2}\right)\left(m_{D}^{2}-p^{2}\right)-p^{2}\, f_{1}^{2}\left(\left|y_{D}\right|^{2}\left(m_{U}^{2}-p^{2}\right)+\left|y_{U}\right|^{2}\left(m_{D}^{2}-p^{2}\right)\right)},
\end{align}
\begin{align}
\Pi_{10}^{u}\left(p^{2}\right) & =\frac{f_{0}^{2}\left|\lambda_{u}\right|^{2}}{m_{U}^{2}-p^{2}},\\
\Pi_{\overline{10}}^{u}\left(p^{2}\right) & =\frac{f_{0}^{2}\left|\lambda_{u}\right|^{2}}{m_{U}^{2}-p^{2}},\\
\Pi_{15}^{u}\left(p^{2}\right) & =\frac{f_{0}^{2}\left|\lambda_{u}\right|^{2}\left(-f_{1}^{2}p^{2}\left|y_{D}\right|^{2}+f_{1}^{2}\left|y_{U}\right|^{2}\left(m_{D}^{2}-p^{2}\right)+\left(m_{D}^{2}-p^{2}\right)\left(m_{Q}^{2}-p^{2}\right)\right)}{\left(m_{Q}^{2}-p^{2}\right)\left(m_{U}^{2}-p^{2}\right)\left(m_{D}^{2}-p^{2}\right)-p^{2}\, f_{1}^{2}\left(\left|y_{D}\right|^{2}\left(m_{U}^{2}-p^{2}\right)+\left|y_{U}\right|^{2}\left(m_{D}^{2}-p^{2}\right)\right)},
\end{align}
\begin{align}
\Pi_{10}^{d}\left(p^{2}\right) & =\frac{f_{0}^{2}\left|\lambda_{d}\right|^{2}}{m_{D}^{2}-p^{2}},\\
\Pi_{\overline{10}}^{d}\left(p^{2}\right) & =\frac{f_{0}^{2}\left|\lambda_{d}\right|^{2}}{m_{D}^{2}-p^{2}},\\
\Pi_{15}^{d}\left(p^{2}\right) & =\frac{f_{0}^{2}\left|\lambda_{d}\right|^{2}\left(-f_{1}^{2}p^{2}\left|y_{U}\right|^{2}+f_{1}^{2}\left|y_{D}\right|^{2}\left(m_{U}^{2}-p^{2}\right)+\left(m_{U}^{2}-p^{2}\right)\left(m_{Q}^{2}-p^{2}\right)\right)}{\left(m_{Q}^{2}-p^{2}\right)\left(m_{U}^{2}-p^{2}\right)\left(m_{D}^{2}-p^{2}\right)-p^{2}\, f_{1}^{2}\left(\left|y_{D}\right|^{2}\left(m_{U}^{2}-p^{2}\right)+\left|y_{U}\right|^{2}\left(m_{D}^{2}-p^{2}\right)\right)},\\
M_{15}^{u}\left(p^{2}\right) & =\frac{f_{0}^{2}\, f_{1}\, y_{U}\,\lambda_{u}^{*}\,\lambda_{q}\, m_{Q}\, m_{U}\left(m_{D}^{2}-p^{2}\right)}{\left(m_{Q}^{2}-p^{2}\right)\left(m_{U}^{2}-p^{2}\right)\left(m_{D}^{2}-p^{2}\right)-p^{2}\, f_{1}^{2}\left(\left|y_{D}\right|^{2}\left(m_{U}^{2}-p^{2}\right)+\left|y_{U}\right|^{2}\left(m_{D}^{2}-p^{2}\right)\right)},\\
M_{15}^{d}\left(p^{2}\right) & =\frac{f_{0}^{2}\, f_{1}\, y_{D}\,\lambda_{d}^{*}\,\lambda_{q}\, m_{Q}\, m_{D}\left(m_{U}^{2}-p^{2}\right)}{\left(m_{Q}^{2}-p^{2}\right)\left(m_{U}^{2}-p^{2}\right)\left(m_{D}^{2}-p^{2}\right)-p^{2}\, f_{1}^{2}\left(\left|y_{D}\right|^{2}\left(m_{U}^{2}-p^{2}\right)+\left|y_{U}\right|^{2}\left(m_{D}^{2}-p^{2}\right)\right)}.
\end{align}

The boson form-factors are:
\begin{align}
\Pi_{6}\left(p^{2}\right)=&\frac{f_{0}^{2}\left(f_{1}^{2}g_{1}^{2}-2\, p^{2}\right)}{2\left(f_{0}^{2}+f_{1}^{2}\right)g_{1}^{2}-4\, p^{2}},\\
\Pi_{15}\left(p^{2}\right)=&-\frac{f_{0}^{2}p^{2}}{f_{0}^{2}g_{1}^{2}-2p^{2}},
\label{E:FacFormBos}
\end{align}

The case without $d_R$ resonances can be obtained by taking $\lambda_d=0$ and throwing the terms depending on $y_D$ and $m_D$.

\section{Useful equations}\label{ap-tools}
In the present appendix we present some mathematical results used in the paper.

The function resulting from the triangle-loop of fermions, $A_{1/2}$ is defined by:
\begin{align}
&A_{1/2}\left(\tau\right)=\frac{2}{\tau^2}\left[\tau+\left(\tau-1\right)f\left(\tau\right)\right]\ .
\end{align}
If the triangle loop is formed by vector bosons instead of fermions, the loop function is:
\begin{align}
&A_{1}\left(\tau\right)=-\frac{1}{\tau^2}\left[2\tau^2+3\tau+3\left(2\tau-1\right)f\left(\tau\right)\right].
\end{align}
Finally, if the particles in the loop are scalar bosons, the loop function is:
\begin{align}
&A_{0}\left(\tau\right)=-\frac{1}{\tau^2}\left[\tau-f\left(\tau\right)\right].
\end{align}
For the three cases, we define the function $f$ as follows:
\begin{align}
&f\left(\tau\right)=\begin{cases}
\arcsin^{2}\left(\sqrt{\tau}\right) & \tau\leqslant1 \ ,\\
-\frac{1}{4}\left[\log\left(\frac{1+\sqrt{1-\tau^{-1}}}{1-\sqrt{1-\tau^{-1}}}\right)-i\pi\right]^2 & \tau>1 \ .
\end{cases}
\end{align}
These functions are used in the one-loop contribution to the decays of the Higgs boson to gg, $\gamma\gamma$ and Z$\gamma$.
\bibliographystyle{hunsrt}
\bibliography{bibPNGB}

\begin{thebibliography}{10}

\bibitem{Grojean:2013qca}
Christophe Grojean, Oleksii Matsedonskyi, and Giuliano Panico.
\newblock {Light top partners and precision physics}.
\newblock {\em JHEP}, 10:160, 2013, 1306.4655.

\bibitem{Maldacena:1997re}
Juan~Martin Maldacena.
\newblock {The Large N limit of superconformal field theories and
  supergravity}.
\newblock {\em Int. J. Theor. Phys.}, 38:1113--1133, 1999, hep-th/9711200.
\newblock [Adv. Theor. Math. Phys.2,231(1998)].

\bibitem{Randall:1999ee}
Lisa Randall and Raman Sundrum.
\newblock {A Large mass hierarchy from a small extra dimension}.
\newblock {\em Phys. Rev. Lett.}, 83:3370--3373, 1999, hep-ph/9905221.

\bibitem{ArkaniHamed:2001nc}
Nima Arkani-Hamed, Andrew~G. Cohen, and Howard Georgi.
\newblock {Electroweak symmetry breaking from dimensional deconstruction}.
\newblock {\em Phys. Lett.}, B513:232--240, 2001, hep-ph/0105239.

\bibitem{Agashe:2004rs}
Kaustubh Agashe, Roberto Contino, and Alex Pomarol.
\newblock {The Minimal composite Higgs model}.
\newblock {\em Nucl. Phys.}, B719:165--187, 2005, hep-ph/0412089.

\bibitem{Gripaios:2009pe}
Ben Gripaios, Alex Pomarol, Francesco Riva, and Javi Serra.
\newblock {Beyond the Minimal Composite Higgs Model}.
\newblock {\em JHEP}, 04:070, 2009, 0902.1483.

\bibitem{Mrazek:2011iu}
J.~Mrazek, A.~Pomarol, R.~Rattazzi, M.~Redi, J.~Serra, and A.~Wulzer.
\newblock {The Other Natural Two Higgs Doublet Model}.
\newblock {\em Nucl. Phys.}, B853:1--48, 2011, 1105.5403.

\bibitem{Gripaios:2009dq}
Ben Gripaios.
\newblock {Composite Leptoquarks at the LHC}.
\newblock {\em JHEP}, 02:045, 2010, 0910.1789.

\bibitem{Frigerio:2011zg}
Michele Frigerio, Javi Serra, and Alvise Varagnolo.
\newblock {Composite GUTs: models and expectations at the LHC}.
\newblock {\em JHEP}, 06:029, 2011, 1103.2997.

\bibitem{DeLuca:2018mzn}
Valerio De~Luca, Andrea Mitridate, Michele Redi, Juri Smirnov, and Alessandro
  Strumia.
\newblock {Colored Dark Matter}.
\newblock {\em Phys. Rev.}, D97(11):115024, 2018, 1801.01135.

\bibitem{Gross:2018zha}
Christian Gross, Andrea Mitridate, Michele Redi, Alessandro Strumia, and Juri
  Smirnov.
\newblock {Cosmological Abundance of Colored Relics}.
\newblock 2018, 1811.08418.

\bibitem{Balkin:2017aep}
Reuven Balkin, Maximilian Ruhdorfer, Ennio Salvioni, and Andreas Weiler.
\newblock {Charged Composite Scalar Dark Matter}.
\newblock {\em JHEP}, 11:094, 2017, 1707.07685.

\bibitem{Chala:2016ykx}
Mikael Chala, Germano Nardini, and Ivan Sobolev.
\newblock {Unified explanation for dark matter and electroweak baryogenesis
  with direct detection and gravitational wave signatures}.
\newblock {\em Phys. Rev.}, D94(5):055006, 2016, 1605.08663.

\bibitem{Balkin:2018tma}
Reuven Balkin, Maximilian Ruhdorfer, Ennio Salvioni, and Andreas Weiler.
\newblock {Dark matter shifts away from direct detection}.
\newblock {\em JCAP}, 1811(11):050, 2018, 1809.09106.

\bibitem{Contino:2004vy}
Roberto Contino and Alex Pomarol.
\newblock {Holography for fermions}.
\newblock {\em JHEP}, 11:058, 2004, hep-th/0406257.

\bibitem{Agashe:2004cp}
Kaustubh Agashe, Gilad Perez, and Amarjit Soni.
\newblock {Flavor structure of warped extra dimension models}.
\newblock {\em Phys. Rev.}, D71:016002, 2005, hep-ph/0408134.

\bibitem{Contino:2003ve}
Roberto Contino, Yasunori Nomura, and Alex Pomarol.
\newblock {Higgs as a holographic pseudoGoldstone boson}.
\newblock {\em Nucl. Phys.}, B671:148--174, 2003, hep-ph/0306259.

\bibitem{Contino:2006nn}
Roberto Contino, Thomas Kramer, Minho Son, and Raman Sundrum.
\newblock {Warped/composite phenomenology simplified}.
\newblock {\em JHEP}, 05:074, 2007, hep-ph/0612180.

\bibitem{Carena:2014ria}
Marcela Carena, Leandro Da~Rold, and Eduardo Pont\'on.
\newblock {Minimal Composite Higgs Models at the LHC}.
\newblock {\em JHEP}, 06:159, 2014, 1402.2987.

\bibitem{Agashe:2006at}
Kaustubh Agashe, Roberto Contino, Leandro Da~Rold, and Alex Pomarol.
\newblock {A Custodial symmetry for $Zb \bar b$}.
\newblock {\em Phys. Lett.}, B641:62--66, 2006, hep-ph/0605341.

\bibitem{Csaki:2008zd}
Csaba Csaki, Adam Falkowski, and Andreas Weiler.
\newblock {The Flavor of the Composite Pseudo-Goldstone Higgs}.
\newblock {\em JHEP}, 09:008, 2008, 0804.1954.

\bibitem{Agashe:2008uz}
Kaustubh Agashe, Aleksandr Azatov, and Lijun Zhu.
\newblock {Flavor Violation Tests of Warped/Composite SM in the Two-Site
  Approach}.
\newblock {\em Phys. Rev.}, D79:056006, 2009, 0810.1016.

\bibitem{Coleman:1969sm}
Sidney~R. Coleman, J.~Wess, and Bruno Zumino.
\newblock {Structure of phenomenological Lagrangians. 1.}
\newblock {\em Phys. Rev.}, 177:2239--2247, 1969.

\bibitem{Callan:1969sn}
Curtis~G. Callan, Sidney~R. Coleman, J.~Wess, and Bruno Zumino.
\newblock {Structure of phenomenological Lagrangians. 2.}
\newblock {\em Phys. Rev.}, 177:2247--2250, 1969.

\bibitem{Azatov:2011qy}
Aleksandr Azatov and Jamison Galloway.
\newblock {Light Custodians and Higgs Physics in Composite Models}.
\newblock {\em Phys. Rev.}, D85:055013, 2012, 1110.5646.

\bibitem{Falkowski:2007hz}
Adam Falkowski.
\newblock {Pseudo-goldstone Higgs production via gluon fusion}.
\newblock {\em Phys. Rev.}, D77:055018, 2008, 0711.0828.

\bibitem{Coleman:1973jx}
Sidney~R. Coleman and Erick~J. Weinberg.
\newblock {Radiative Corrections as the Origin of Spontaneous Symmetry
  Breaking}.
\newblock {\em Phys. Rev.}, D7:1888--1910, 1973.

\bibitem{Panico:2011pw}
Giuliano Panico and Andrea Wulzer.
\newblock {The Discrete Composite Higgs Model}.
\newblock {\em JHEP}, 09:135, 2011, 1106.2719.

\bibitem{Panico:2012uw}
Giuliano Panico, Michele Redi, Andrea Tesi, and Andrea Wulzer.
\newblock {On the Tuning and the Mass of the Composite Higgs}.
\newblock {\em JHEP}, 03:051, 2013, 1210.7114.

\bibitem{DaRold:2018moy}
Leandro Da~Rold and Federico Lamagna.
\newblock {Composite Higgs and leptoquarks from a simple group}.
\newblock 2018, 1812.08678.

\bibitem{Csaki:2018zzf}
Csaba Csáki, Teng Ma, Jing Shu, and Jiang-Hao Yu.
\newblock {Emergence of Maximal Symmetry}.
\newblock 2018, 1810.07704.

\bibitem{Chala:2018opy}
Mikael Chala, Maria Ramos, and Michael Spannowsky.
\newblock {Gravitational wave and collider probes of extended Higgs sectors
  with a low cutoff}.
\newblock 2018, 1812.01901.

\bibitem{Sirunyan:2018fki}
Albert~M Sirunyan et~al.
\newblock {Search for a W' boson decaying to a vector-like quark and a top or
  bottom quark in the all-jets final state}.
\newblock {\em Submitted to: JHEP}, 2018, 1811.07010.

\bibitem{Sirunyan:2018rfo}
Albert~M Sirunyan et~al.
\newblock {Search for a heavy resonance decaying to a top quark and a
  vector-like top quark in the lepton+jets final state in pp collisions at
  $\sqrt{s} =$ 13 TeV}.
\newblock 2018, 1812.06489.

\bibitem{Aaboud:2018uek}
Morad Aaboud et~al.
\newblock {Search for pair production of heavy vector-like quarks decaying into
  high-$p_T$ $W$ bosons and top quarks in the lepton-plus-jets final state in
  $pp$ collisions at $\sqrt{s}=13$ TeV with the ATLAS detector}.
\newblock {\em JHEP}, 08:048, 2018, 1806.01762.

\bibitem{Aaboud:2018xuw}
Morad Aaboud et~al.
\newblock {Search for pair production of up-type vector-like quarks and for
  four-top-quark events in final states with multiple $b$-jets with the ATLAS
  detector}.
\newblock {\em JHEP}, 07:089, 2018, 1803.09678.

\bibitem{ATLAS:2018yey}
The~ATLAS collaboration.
\newblock {Reinterpretation of searches for supersymmetry in models with
  variable $R$-parity-violating coupling strength and long-lived $R$-hadrons}.
\newblock 2018.

\bibitem{CMS:2016ybj}
CMS Collaboration.
\newblock {Search for heavy stable charged particles with
  $12.9~\mathrm{fb}^{-1}$ of 2016 data}.
\newblock 2016.

\bibitem{Carena:2012xa}
Marcela Carena, Ian Low, and Carlos E.~M. Wagner.
\newblock {Implications of a Modified Higgs to Diphoton Decay Width}.
\newblock {\em JHEP}, 08:060, 2012, 1206.1082.

\bibitem{ATLAS:2012klq}
Georges Aad et~al.
\newblock {Measurements of the Higgs boson production and decay rates and
  coupling strengths using pp collision data at $\sqrt{s}=7$ and 8 TeV in the
  ATLAS experiment}.
\newblock {\em Eur. Phys. J.}, C76(1):6, 2016, 1507.04548.

\bibitem{ATLAS:2013oma}
{Measurements of the properties of the Higgs-like boson in the two photon decay
  channel with the ATLAS detector using 25 $\mathrm{fb}^{-1}$ of proton-proton
  collision data}.
\newblock 2013.

\bibitem{Chatrchyan:2013mxa}
Serguei Chatrchyan et~al.
\newblock {Measurement of the properties of a Higgs boson in the four-lepton
  final state}.
\newblock {\em Phys. Rev.}, D89(9):092007, 2014, 1312.5353.

\bibitem{Konig:2014iqa}
Matthias K{\" o}nig, Matthias Neubert, and David~M. Straub.
\newblock {Dipole operator constraints on composite Higgs models}.
\newblock {\em Eur. Phys. J.}, C74(7):2945, 2014, 1403.2756.

\bibitem{Antipin:2014mda}
Oleg Antipin, Stefania De~Curtis, Michele Redi, and Carlotta Sacco.
\newblock {Muon magnetic moment and the pseudo-Goldstone Higgs boson}.
\newblock {\em Phys. Rev.}, D90(6):065016, 2014, 1407.2471.

\bibitem{Panico:2018hal}
Giuliano Panico, Alex Pomarol, and Marc Riembau.
\newblock {EFT approach to the electron Electric Dipole Moment at the two-loop
  level}.
\newblock 2018, 1810.09413.

\bibitem{Andreev:2018ayy}
V.~Andreev et~al.
\newblock {Improved limit on the electric dipole moment of the electron}.
\newblock {\em Nature}, 562(7727):355--360, 2018.

\end{thebibliography}

\end{document}